\documentclass[onecolumn,draftclsnofoot]{IEEEtran} 
\usepackage{tikz}
\usepackage{amsmath,amssymb,bm,amsthm}
\usepackage{graphicx}
\usepackage{soul} 
\usepackage{diagbox}  
\usepackage{cite}
\usepackage{hyperref} 
\usepackage{cleveref}
\usepackage{enumerate}
\usepackage{url}
\usepackage{balance}   
\allowdisplaybreaks[2]  
\usepackage{pdflscape}
\usepackage{array}
\newcolumntype{C}[1]{>{\centering\let\newline\\\arraybackslash\hspace{0pt}}m{#1}} 

\begin{document}
	\title{Sliding Secure Symmetric Multilevel Diversity Coding}   
    \author{\IEEEauthorblockN{Tao Guo, Laigang Guo, Yinfei Xu, Congduan Li, Shi Jin, Raymond Yeung}
	}
	\maketitle
	\balance  
	\newcommand{\reffig}[1]{Fig. \ref{#1}}
	\newcommand{\cA}{\mathcal{A}}
	\newcommand{\cB}{\mathcal{B}}
	\newcommand{\cD}{\mathcal{D}}
	\newcommand{\cK}{\mathcal{K}}
	\newcommand{\cL}{\mathcal{L}}
	\newcommand{\cO}{\mathcal{O}}
	\newcommand{\cR}{\mathcal{R}}
	\newcommand{\cT}{\mathcal{T}}
	\newcommand{\cU}{\mathcal{U}}
	\newcommand{\cX}{\mathcal{X}}

        \theoremstyle{plain}
	\newtheorem{theorem}{Theorem}
	\newtheorem{lemma}{Lemma}
	\newtheorem{corollary}{Corollary}[theorem]
	\newtheorem{proposition}{Proposition}
	\newtheorem{conjecture}{Conjecture}
	\newtheorem{claim}{Claim}

\newtheorem*{namedthm*}{\thistheoremname}
\newcommand{\thistheoremname}{} 
\newenvironment{namedthm}[1]
{\renewcommand{\thistheoremname}{#1}\begin{namedthm*}}
	{\end{namedthm*}}
	
	\theoremstyle{remark}
	\newtheorem{remark}{Remark}
	
	\theoremstyle{definition}
	\newtheorem{definition}{Definition}
	\newtheorem{example}{Example}

\begin{abstract}
	Symmetric multilevel diversity coding (SMDC) is a source coding problem 
	where the independent sources are ordered according to their importance. 
	It was shown that separately encoding independent sources (referred to as ``\textit{superposition coding}") is optimal. 
	In this paper, we consider an $(L,s)$ \textit{sliding secure} SMDC problem with security priority, 
	where each source $\bm{X}_{\alpha}~(s\leq \alpha\leq L)$ is kept perfectly secure if no more than $\alpha-s$ encoders are accessible. 
	The reconstruction requirements of the $L$ sources are the same as classical SMDC. 
	A special case of an $(L,s)$ sliding secure SMDC problem that the first $s-1$ sources are constants 
	is called the $(L,s)$ \textit{multilevel secret sharing} problem. 
	For $s=1$, the two problems coincide, and we show that superposition coding is optimal. 
	The rate regions for the $(3,2)$ problems are characterized. 
	It is shown that superposition coding is suboptimal for both problems. 
	The main idea that joint encoding can reduce coding rates is that 
	we can use the previous source $\bm{X}_{\alpha-1}$ as the secret key of~$\bm{X}_{\alpha}$. 
	Based on this idea, we propose a coding scheme that achieves the 
	minimum sum rate of the general $(L,s)$ multilevel secret sharing problem. 
	Moreover, superposition coding of the $s$ sets of sources 
	$\bm{X}_1$, $\bm{X}_2$, $\cdots$, $\bm{X}_{s-1}$, $(\bm{X}_s, \bm{X}_{s+1}, \cdots, \bm{X}_L)$ 
	achieves the minimum sum rate of the general sliding secure SMDC problem.
\end{abstract}
\section{Introduction}
Symmetric multilevel diversity coding (SMDC) is a distributed source coding problem 
where there are $L$ independent discrete memoryless sources $\{X_1(t),X_2(t),\cdots,X_L(t)\}_{t=1}^{\infty}$ with time index $t$. 
The sources are encoded by $L$ encoders. The decoder can access a subset $\cU$ of the encoders. 
The first $\alpha$ sources $X_1(t),X_2(t),\cdots,X_\alpha(t)$ are required to be 
reconstructed losslessly at the decoder if any $|\cU|\geq \alpha$ for any $\alpha=1,2,\cdots,L$. 
The word ``symmetric" refers to the fact that the sources that are required to be 
losslessly reconstructed depend on the subset of accessible encoders  only through its cardinality. 
A simple coding strategy, referred to as \textit{superposition coding} \cite{yeung97}\cite{yeung99}, 
is to encode the independent sources separately. 
The optimality of superposition coding was established by Yeung and Zhang \cite{yeung99}, 
where they provided an implicit description of the superposition region. 
An explicit characterization of the rate constraints was achieved in \cite{SMDC-GuoYeung-IT20}.
The problem is also known as priority encoding transmission, which was independently studied in \cite{SMDC-PET-IT96}.
The SMDC problem has been extended, to the distributed setting~\cite{ChenJun-distributedSMDC-IT15}, to allow node regeneration~\cite{Tian-SMDC-reg-IT16,Shao-secureReg-Entropy18}, to allow asymmetry~\cite{mohajer-tian-diggavi10,Li-SMDC-compute-IT17,Li-asymmetric-secure-JSAC18,Li-asymmetric-perfectSecure-TVT17}, and to involve security constraints~\cite{Li-asymmetric-secure-JSAC18,Li-asymmetric-perfectSecure-TVT17,liutie13-sSMDC,WS-SMDC-IT20}.

Secret sharing is a closely related problem, where a secret is distributed among a group of participants. 
In the $(k,L)$-\textit{threshold secret sharing} problem \cite{Shamir79}, 
the \textit{secret} (referred to as the ``source" hereafter) is encoded into $L$ messages. 
The source can be recovered losslessly from any $k$ of them while 
it is kept perfectly secure if less than $k$ messages are available. 
If we relax the security constraints so that any $c~(0\leq c<k)$ messages provide no information about the source, 
then the problem becomes the $(c,k,L)$-\textit{ramp secret sharing} problem \cite{yamamoto85}\cite{blakley85-ramp-secret-sharing}. 
The optimality of ramp secret sharing schemes require that the sum of any $k-c$ rates is 
greater than or equal to the entropy of the source \cite{ramp-secret-sharing96}\cite{liutie13-sSMDC}. 
The rate region depends only on the difference $k-c$ rather than specific $k$ and $c$. 
The difference $k-c$ is the gap between perfect secrecy and losslessly reconstruction, which is an important concept in the sequel.

The problem of SMDC was generalized to the weakly secure setting in \cite{WS-SMDC-IT20}, which characterized the conditions that superposition coding is sum-rate optimal. 
The rate region was further obtained for a special case where a set of more important messages are maximally secure 
and other less important messages do not have any security guarantee.
This results in extreme security conditions that some sources are protected to the best and not easy to recover while others are not protected at all. 
However, in some circumstance, more flexible security are required. 
In the current paper, we reformulate the security in a different manner which is linear with $\alpha$. 

Moreover, all the previous work focused on the recovery priority, i.e., more important sources should be recovered prior to the less important ones, which means more important sources are less secure since they are easier to recover. 
In contrast, we consider security priority, where a more important source should be more secure than less important ones. 
This may result in a recovery order reverse to the classical SMDC that the recovery of more important sources requires access to more encoders. 
For notational consistence, note that we assume the importance of the sources $\{X_l(t)\}$ increases with the subscript $l$ in the sequel, which is a reverse order compared to the previous work.

Specifically, we consider the problem of \textit{sliding secure} SMDC with security priority (abbr. sliding secure SMDC), 
where the word ``sliding" refers to the feature that the security constraints vary linearly with the source subscripts. 
For $1\leq s\leq L$, the $(L,s)$ sliding secure SMDC problem consists of $L$ 
independent discrete memoryless sources $\{X_1(t),X_2(t),\cdots,X_L(t)\}_{t=1}^{\infty}$, which are encoded by a set of $L$ encoders indexed by $\cL=\{1,2,\cdots,L\}$. 
A decoder and an eavesdropper have access to a subset $\cU\subseteq\cL$ and $\cA\subseteq\cL$ of encoders, respectively. 
For $1\leq \alpha\leq L$, the decoder is required to losslessly 
reconstruct the first $\alpha$ sources if it can access any $\alpha$ encoders. 
For $s\leq \alpha\leq L$, source $X_{\alpha}(t)$ should be kept 
perfectly secure from the eavesdropper if no more than $\alpha-s$ encoders are accessible. 
The parameter $s$ is the gap between perfect secrecy and lossless reconstruction for each of the sources $X_{s}(t),X_{s+1}(t),\cdots,X_L(t)$. 
There is no security constraints for the first $s-1$ sources. 
The word ``symmetric", similar as in classical SMDC, refers to the fact that security and reconstruction 
depend on the subset of accessible encoders only via its cardinality rather than which specific subset is accessible.  
The $(L,s)$ sliding secure SMDC problem reduces to the classical SMDC problem when $s=L$.

If the first $s-1$ sources are constants, the problem reduces to the $(L,s)$ \textit{multilevel secret sharing} problem. 
More specifically, we only have sources $X_{s}(t),X_{s+1}(t),\cdots,X_L(t)$. 
For $s\leq \alpha\leq L$, source $X_{\alpha}(t)$ should be losslessly reconstructed if the decoder have access to any $\alpha$ encoders 
and be kept perfectly secure if the eavesdropper can only access no more than $\alpha-s$ encoders.

As mentioned, superposition coding is optimal for SMDC. 
One may ask whether superposition coding is also optimal for the multilevel secret sharing and sliding secure SMDC problems. 
For $s=1$, the answer is ``yes". 
However, for $s\geq 2$, superposition of the independent sources is suboptimal. 
The main idea that joint encoding can reduce coding rates is that 
we can use the previous source $X_{\alpha-1}(t)$ as the secret keys for the source $X_{\alpha}(t)$ for $s+1\leq \alpha\leq L$. 
Based on this idea, we propose a coding scheme for the $(3,2)$ multilevel secret  sharing problem that achieves the entire rate region. 
For the $(3,2)$ sliding secure SMDC problem, superposition of two sets of sources $X_1(t)$ and $(X_2(t),X_3(t))$ is shown to be optimal.

For any $s\geq 2$, we propose a coding scheme that achieves the minimum sum rate of the multilevel secret sharing problem. 
We also show that superposition of the $s$ sets of sources 
$X_1(t)$, $X_2(t)$, $\cdots$, $X_{s-1}(t)$, $(X_{s}(t),X_{s+1}(t), \cdots, X_L(t))$ 
achieves the minimum sum rate of the sliding secure SMDC problem. 
However, it is difficult to determine whether such coding schemes can achieve 
the entire rate region of the respective problems since it is really challenging to fully characterize the regions. 
We may pay more attention to this direction in the future work.

Our main contributions are summarized as follows: 
\begin{enumerate}
    \item We propose the $(L,s)$ sliding secure SMDC problem and multilevel secret sharing problem.

    \item The optimal rate regions of both problems are fully characterized for $(L,s)=(3,2)$, which shows the suboptimality of superposition coding. A joint coding scheme is proposed to achieve the rate region, where $X_2(t)$ is used as the secret key for $X_3(t)$ so that coding rates are reduced. 
    
    \item A pseudo superposition coding scheme is proposed to achieve the minimum sum rate of the general sliding secure SMDC problem, which uses superposition for the $s$ sets of sources 
    $X_1(t)$, $X_2(t)$, $\cdots$, $X_{s-1}(t)$, $(X_{s}(t),X_{s+1}(t), \cdots, X_L(t))$ and joint encoding among $X_s(t), X_{s+1}(t), \cdots, X_L(t)$.
\end{enumerate}

The rest of the paper is organized as follows. 
We describe the sliding secure SMDC and multilevel secret sharing problems in \Cref{section-formualtion}. 
For $s=1$, we provide a simple proof of the optimality of superposition in \Cref{section-L1}. 
The entire rate region of the $(3,2)$ multilevel secret sharing and sliding secure SMDC problems are characterized in \Cref{section-32mss,section-32sMDC}. 
In \Cref{section-Ls-mss}, we propose a general coding scheme for 
the $(L,s)$ multilevel secret sharing problem that achieves the minimum sum rate for $s\geq 2$. 
A coding scheme that achieves the minimum sum rate of the sliding secure SMDC problem is given in \Cref{section-Ls-sMDC}. 
We conclude that paper in \Cref{section-conclusion}. Some essential proofs can be found in the appendices.

\section{Problem Formulation} \label{section-formualtion}
\subsection{Sliding Secure SMDC}
\begin{figure}[!t]
	\centering
	\begin{tikzpicture}[font=\footnotesize]
	
	\draw [->,>=stealth](-0.7,0)--(-0.2,0);
    \node (source) at (-0.9,0.4) {$\bm{X}_1$};
    \node at (-0.9,0) {$\vdots$};
    \node at (-0.9,-0.5) {$\bm{X}_L$};
	\draw [thick] (-0.2,-1.0) rectangle (1.0,1.0) node(encoder)[midway] {Encoder};	
	
	\draw[thick] (3.5,-1.0) rectangle (4.7,1.0) node (decoder)[midway]{Decoder};
	\draw [->,>=stealth] (4.7,0)--(5.0,0);
	\draw (5.0,0) node [right]{$\bm{X}_1,\bm{X}_2,\cdots,\bm{X}_{\alpha}$};
	
	\draw (1.0,0.6)--(2.6,0.6)--(3.0,0.8); \draw (3.0,0.6)--(3.5,0.6); \node at (1.25,0.75) {$W_1$};
	\draw (1.0,0.2)--(2.6,0.2)--(3.0,0.4); \draw (3.0,0.2)--(3.5,0.2); \node at (1.25,0.35) {$W_2$};
	\draw (1.25,0) node {$\vdots$};
	\draw  (1.0,-0.7)--(2.6,-0.7)--(3.0,-0.5); \draw (3.0,-0.7)--(3.5,-0.7); \node at (1.25,-0.52) {$W_L$};
	
	\draw[thick] (3.5,-1.3) rectangle (4.7,-3.3); \node at (4.1,-2.1){Eaves-};\node at (4.1,-2.4) {dropper};
	\draw [->,>=stealth] (4.7,-2.3)--(5.0,-2.3);
	\draw (5.0,-2.3) node [right]{no information of $\bm{X}_{\alpha}$};
	
	\draw (2.2,0.6)--(2.2,-1.7)--(2.6,-1.7)--(3.0,-1.5); \draw (3.0,-1.7)--(3.5,-1.7);
	\draw (1.8,0.2)--(1.8,-2.1)--(2.6,-2.1)--(3.0,-1.9); \draw (3.0,-2.1)--(3.5,-2.1);
	\draw (2.5,-2.5) node {$\vdots$};
	\draw  (1.3,-0.7)--(1.3,-3.0)--(2.6,-3.0)--(3.0,-2.8); \draw (3.0,-3.0)--(3.5,-3.0);
	\fill [black] (2.2,0.6) circle (1.3pt); \fill [black] (1.8,0.2) circle (1.3pt); \fill [black] (1.3,-0.7) circle (1.3pt);

        \draw[dashed] (2.8,-2.35) ellipse (0.5 and 1.1);
	\node at (2.8,-3.7) {$|\cA|\leq [\alpha-s]^+$};
	\draw[dashed] (2.8,0.1) ellipse (0.5 and 1.0);
	\node at (2.8,1.25) {$|\cU|=\alpha$};
	\end{tikzpicture}
	\caption{The sliding secure SMDC model.}
	\label{fig_sliding secure SMDC}
\end{figure}
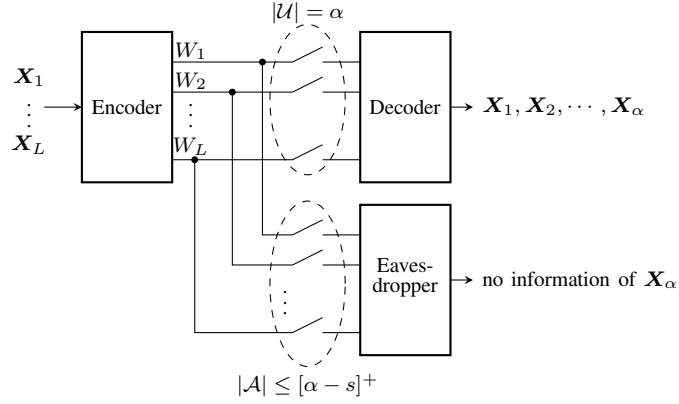

Let $t$ be a time index and $\{X_1(t),X_2(t),\cdots,X_L(t)\}_{t=1}^{\infty}$ be a collection of $L$ independent discrete memoryless sources with an $L$-tuple of generic random variables $(X_1,X_2,\cdots,X_L)$ taking values in the finite alphabets $\cX_1\times\cX_2\times\cdots\times\cX_L$. In the sequel, we use boldface letters to denote vectors of length $n$, for example $\bm{X}_1=(X_1[1],X_1[2],\cdots,X_1[n])$. For simplicity, assume all the entropies in this paper are in the unit of bit. The sliding secure SMDC problem is depicted in \reffig{fig_sliding secure SMDC}. 
For $1\leq s\leq L$, the $(L,s)$ sliding secure SMDC problem consists of a set of $L$ encoders indexed by $\cL=\{1,2,\cdots,L\}$, a decoder that accesses a subset $\cU\subseteq\cL$ of encoders, and an eavesdropper who has access to a subset $\cA\subseteq\cL$ of encoders. Each of the encoders has access to all the sources. For any $x\in\mathbb{R}$, define a function by
\begin{equation}
[x]^+=\begin{cases}x,& \text{ if }x>0\\0,&\text{ if }x\leq 0.\end{cases}
\end{equation}
For $1\leq \alpha\leq L$, no matter which specific subset of encoders are accessible, the decoder should be able to reconstruct the first $\alpha$ sources if $|\cU|=\alpha$ and the source $\bm{X}_{\alpha}$ should be kept perfectly secure from the eavesdropper if $|\cA|\leq[\alpha-s]^+$.

Let $K$ be a secret key taking values in some finite key space $\cK$. Note that we don't limit the size of $\cK$ in this paper. The secret key is shared by all the encoders but not the decoder or the eavesdropper. An $(n,M_1,M_2,\cdots,M_L)$ code is defined by $L$ encoding functions
\begin{equation}
f_l:\prod_{i=1}^L\cX_i^n\times\cK\rightarrow\{1,2,\cdots,M_l\}, \text{ for }l\in\cL \label{encoding fn}
\end{equation}
and decoding functions
\begin{equation}
g_{_{\cU}}:\prod_{l\in\cU}\{1,2,\cdots,M_l\}\rightarrow\prod_{i=1}^{\alpha}\cX_i^n,\text{ for }\cU\subseteq\cL\text{ s.t. }|\cU|=\alpha,  \label{decoding fn}
\end{equation}
for $1\leq\alpha\leq L$. Let
\begin{equation}
W_l\triangleq f_l(\bm{X}_1,\bm{X}_2,\cdots,\bm{X}_L,K)
\end{equation}
be the output of Encoder-$l$. For any $\cB\subseteq\cL$, let $W_{\cB}=\{W_l:l\in\cB\}$. A nonnegative rate tuple $(R_1,R_2,\cdots,R_L)$ is \textit{admissible} for the $(L,s)$ sliding secure SMDC problem if for any $\epsilon>0$, there exists, for sufficiently large $n$, an $(n,M_1,M_2,\cdots,M_L)$ code such that
\begin{equation}
\frac{1}{n}\log M_l\leq R_l+\epsilon,\forall~ l\in\cL, \label{rate def}
\end{equation}
and for all $\alpha=1,2,\cdots,L$,
\begin{align}
g_{_{\cU}}(W_{\cU})=\bm{X}_1,\bm{X}_2,\cdots,\bm{X}_{\alpha}, &~\forall~ \cU\subseteq\cL\text{ s.t. }|\cU|=\alpha,  \label{recover constraint}\\
H(\bm{X}_{\alpha}|W_{\cA})=H(\bm{X}_{\alpha}),&~\forall~\cA\subseteq\cL\text{ s.t. }|\cA|\leq [\alpha-s]^+. \label{security constraint}
\end{align}
The admissible rate region $\cR_{L,s}$ is the collection of all admissible rate tuples.

When $s=L$, there are no security constraints,  and thus the $(L,s)$ sliding secure SMDC problem becomes the classical $L$-channel SMDC problem.

\subsection{Multilevel Secret Sharing}
The $(L,s)$ sliding secure SMDC problem reduces to the $(L,s)$ multilevel secret sharing problem if the first $s-1$ sources are constants, i.e. 
\begin{equation}
H(X_1)=H(X_2)=\cdots=H(X_{s-1})=0.
\end{equation}
Specifically, a collection of $L-s+1$ independent discrete memoryless sources $\bm{X}_s,\bm{X}_{s+1},\cdots,\bm{X}_L$ are encoded by $L$ encoders indexed by $\cL$. A decoder and an eavesdropper have access to a subset of $\cU\subseteq\cL$ and $\cA\subseteq\cL$ of encoders, respectively. For $s\leq \alpha\leq L$, the decoder is required to reconstruct sources $\bm{X}_s,\bm{X}_{s+1},\cdots,\bm{X}_{\alpha}$ if $|\cU|=\alpha$ and the source $\bm{X}_{\alpha}$ should be kept perfectly secure from the eavesdropper if $|\cA|\leq \alpha-s$. The $(n,M_1,M_2,\cdots,M_L)$ code reduces to the encoding functions
\begin{equation}
f_l:\prod_{i=s}^L\cX_i^n\times\cK\rightarrow\{1,2,\cdots,M_l\}, \text{ for all }l\in\cL
\end{equation}
and decoding functions
\begin{equation}
g_{_{\cU}}:\prod_{l\in\cU}\{1,2,\cdots,M_l\}\rightarrow\prod_{i=s}^{\alpha}\cX_i^n,\text{ for all }\cU\subseteq\cL\text{ s.t. }|\cU|=\alpha,
\end{equation}
for $s\leq\alpha\leq L$. The output of Encoder-$l$ becomes
\begin{equation}
W_l= f_l(\bm{X}_{s},\bm{X}_{s+1},\cdots,\bm{X}_L,K).
\end{equation}
A nonnegative rate tuple $(R_1,R_2,\cdots,R_L)$ is \textit{admissible} for the $(L,s)$ multilevel secret sharing problem if for any $\epsilon>0$, there exists, for sufficiently large $n$, an $(n,M_1,M_2,\cdots,M_L)$ code such that
\begin{equation}
\frac{1}{n}\log M_l\leq R_l+\epsilon,\forall~ l\in\cL,
\end{equation}
and for all $\alpha=s,s+1,\cdots,L$,
\begin{align}
g_{_{\cU}}(W_{\cU})=\bm{X}_{s},\bm{X}_{s+1},\cdots,\bm{X}_{\alpha},&~\forall~ \cU\subseteq\cL\text{ s.t. }|\cU|=\alpha,  \label{recover constraint-mss} \\
H(\bm{X}_{\alpha}|W_{\cA})=H(\bm{X}_{\alpha}),&~\forall~\cA\subseteq\cL\text{ s.t. }|\cA|\leq \alpha-s. \label{security constraint-mss}
\end{align}
The admissible rate region $\cR_{L,s}^{\text{mss}}$ is the collection of all admissible rate tuples. The relation between sliding secure SMDC and multilevel secret sharing problems implies that
\begin{equation}
\cR_{L,s}^{\text{mss}}=\{(R_1,R_2,\cdots,R_L)\in\cR_{L,s}:H(X_1)=H(X_2)=\cdots=H(X_{s-1})=0\}.
\end{equation}
Denote the rate region of the classical $L$-channel SMDC problem by $\cR_{\text{SMDC}}^L$. Let
\begin{equation}
\cR_{\text{SMDC}}^{L*}=\{(R_1,R_2,\cdots,R_L)\in\cR_{\text{SMDC}}^L:H(X_1)=H(X_2)=\cdots=H(X_{s-1})=0\}.
\end{equation} 

Before the main results, we state some useful existing results about ramp secret sharing in the following subsection.
\subsection{Ramp Secret Sharing}\label{section-rss}
Let $\bm{X}$ be a discrete memoryless source sequence encoded by $L$ encoders. For any $0\leq c<k\leq L$, the $(c,k,L)$ ramp secret sharing problem requires that any subset of no more than $c$ encoders provide no information about the source and any subset of $k$ encoders can losslessly reconstruct the source. The problem is also known as the secure symmetric single-level diversity coding (S-SSDC) in \cite{liutie13-sSMDC}. The admissible rate region $\cR^{\text{rss}}$ is fully characterized in \cite{ramp-secret-sharing96}\cite{liutie13-sSMDC}. The result in summarized is the following lemma. 

Let $\cR(L,k,H)$ be the collection of rate tuples $(R_1,R_2,\cdots,R_L)$ satisfying
\begin{equation}
\sum_{i\in\cB}R_i\geq H
\end{equation}
for any subset $\cB\subseteq\cL$ such that $|\cB|=k$.
\begin{lemma}
	For the $(c,k,L)$ ramp secret sharing problem, the admissible rate region is as follows,
	\begin{equation}
	\cR^{\text{rss}}=\cR(L,k-c,H(X)).  \label{region-rss}
	\end{equation}
\end{lemma}
When $k=c+1$, the $(c,k,L)$ ramp secret sharing problem becomes the $(k,L)$ threshold secret sharing problem and the rate region reduces to $\cR(L,1,H(X))$.


\section{$(L,1)$ sliding secure SMDC}\label{section-L1}
When $s=1$, the $(L,1)$ sliding secure SMDC problem and the $(L,1)$ multilevel secret sharing problem coincide. For $1\leq \alpha\leq L$, the source $\bm{X}_{\alpha}$ can be losslessly reconstructed if the decoder accesses a subset of any $\alpha$ encoders and should be kept perfectly secure if the eavesdropper accesses no more than $\alpha-1$ encoders. The reconstruction and security constraints in \eqref{recover constraint}\eqref{security constraint} become
\begin{align}
H(\bm{X}_{\alpha}|W_{\cU})&=0,~\forall~ \cU\subseteq\cL\text{ s.t. }|\cU|=\alpha,  \label{recover constraint-L,1}\\
H(\bm{X}_{\alpha}|W_{\cA})&=H(\bm{X}_{\alpha}),~\forall~\cA\subseteq\cL\text{ s.t. }|\cA|\leq \alpha-1. \label{security constraint-L,1}
\end{align}
A simple coding scheme for the $(L,1)$ sliding secure SMDC problem is separately encoding the $L$ independent sources, which is referred to as \textit{superposition coding}. 
For each $\alpha\in\cL$, we use the $(\alpha,L)$ threshold secret sharing scheme to encode the single source $\bm{X}_{\alpha}$, which requires a rate of at least $H(X_\alpha)$ at each encoder,i.e., the rate region is simply $\cR\big(L,1,H(X_{\alpha})\big)$.

The superposition region $\cR_{\text{sup}}^1$ induced by separately encoding the $L$ sources is the collection of nonnegative rate tuples $(R_1,R_2,\cdots,R_L)$ such that 
\begin{equation}
R_i=\sum_{\alpha=1}^{L}r_i^{\alpha}, \text{ for }i\in\cL
\end{equation}
where $r_i^{\alpha}\geq 0,~1\leq \alpha\leq L$, and 
\begin{equation}
\big(r_1^{\alpha},r_2^{\alpha},\cdots,r_L^{\alpha}\big)\in\cR\big(L,1,H(X_{\alpha})\big).
\end{equation}
It is easy to eliminate $r_i^{\alpha}~(i,\alpha\in\cL)$ and obtain the following equivalent characterization of the superposition region,
\begin{equation}
\cR_{\text{sup}}^1=\{(R_1,R_2,\cdots,R_L):R_i\geq \sum_{\alpha=1}^{L}H(X_{\alpha}), \text{ for all }i\in\cL\}.  \label{R_sup}
\end{equation}
The following theorem states that superposition coding is optimal for the $(L,1)$ sliding secure SMDC problem and also the $(L,1)$ multilevel secret sharing problem.
\begin{theorem}
	$\cR_{L,1}=\cR_{\text{sup}}^1$. \label{thm-L1}
\end{theorem}
In order to prove \Cref{thm-L1}, we only need to show the converse part.
\begin{proof}
For $1\leq \alpha\leq L$, let $W_1^{\alpha}=(W_1,W_2,\cdots,W_{\alpha})$. From the conditions in \eqref{recover constraint-L,1}\eqref{security constraint-L,1}, we have
\begin{align}
H(\bm{X}_{\alpha})&=H(\bm{X}_{\alpha}|W_2^{\alpha})  \nonumber \\
&=I(\bm{X}_{\alpha};W_{1}|W_2^{\alpha})+H(\bm{X}_{\alpha}|W_{1},W_2^{\alpha})  \nonumber \\
&=I(\bm{X}_{\alpha};W_{1}|W_2^{\alpha})  \nonumber \\
&=H(W_1|W_2^{\alpha})-H(W_1|W_2^{\alpha},\bm{X}_{\alpha}). \label{H(X)-L1}
\end{align}
Thus,
\begin{align}
n\cdot\left(\sum_{\alpha=1}^L H(X_{\alpha})\right)&=\sum_{\alpha=1}^L H(\bm{X}_{\alpha})   \nonumber \\
&= \sum_{\alpha=1}^L\big[H(W_{1}|W_2^{\alpha})-H(W_{1}|W_2^{\alpha},\bm{X}_{\alpha})\big]  \nonumber \\
&=H(W_1)-\sum_{\alpha=1}^{L-1}\big[H(W_1|W_2^{\alpha},\bm{X}_{\alpha})-H(W_1|W_2^{\alpha+1})\big]-H(W_1|W_2^{L},\bm{X}_L)  \nonumber \\
&= H(W_1)-\sum_{\alpha=1}^{L-1}I(W_1;W_{\alpha+1}|W_2^{\alpha},\bm{X}_{\alpha})-H(W_1|W_2^{L},\bm{X}_L)  \label{converse-L_1-R1-3} \\
&\leq H(W_1)  \nonumber \\
&\leq n(R_1+\epsilon)  \label{converse-L_1-R1}
\end{align}
where \eqref{converse-L_1-R1-3} is obtained by
\begin{align}
&\hspace{-0.8cm}H(W_1|W_2^{\alpha},\bm{X}_{\alpha})-H(W_1|W_2^{\alpha+1})  \nonumber \\
&=H(W_1|W_2^{\alpha},\bm{X}_{\alpha})-\big[H(W_1|W_2^{\alpha+1},\bm{X}_{\alpha})+I(W_1;\bm{X}_{\alpha}|W_2^{\alpha+1})\big]  \nonumber \\
&=H(W_1|W_2^{\alpha},\bm{X}_{\alpha})-H(W_1|W_2^{\alpha+1},\bm{X}_{\alpha})  \nonumber \\
&=I(W_1;W_{\alpha+1}|W_2^{\alpha},\bm{X}_{\alpha}) 
\end{align}
for any $1\leq \alpha\leq L-1$. Divide both sides of \eqref{converse-L_1-R1} by $n$ and let $\epsilon\to 0$, we have
\begin{equation}
R_1\geq\sum_{\alpha=1}^{L} H(X_{\alpha}).
\end{equation}
Similarly, for any $i\in\cL$, we can prove that 
\begin{equation}
R_i\geq\sum_{\alpha=1}^{L} H(X_{\alpha}).
\end{equation}
Therefore, \Cref{thm-L1} is proved.
\end{proof}

\begin{remark}
    For $1\leq \alpha_1 < \alpha_2\leq L$, since $\bm{X}_{\alpha_2}$ is always assumed to be more secure than $\bm{X}_{\alpha_1}$, it is impossible to use $\bm{X}_{\alpha_2}$ as a secret key for $\bm{X}_{\alpha_1}$. For $1\leq \alpha\leq L$, if we can access any $\alpha$ encoders, the first $\alpha$ sources $\bm{X}_1,\bm{X}_2,\cdots,\bm{X}_{\alpha}$ are losslessly reconstructed and $\bm{X}_{\alpha+1}$ should be kept perfectly secure, thus it is also impossible to use the first $\alpha$ sources as secret keys for $\bm{X}_{\alpha+1}$ and the successive sources after $\bm{X}_{\alpha+1}$. The fact that any source can not be used as secret keys for other sources provides an intuition of why superposition is optimal for the $(L,1)$ multilevel secret sharing problem.
\end{remark}
\begin{remark}
    For $s=1$, the security constraint \eqref{security constraint} reduces to 
    \begin{equation}
    H(\bm{X}_{\alpha}|W_{\cA})= H(\bm{X}_{\alpha}),~\forall~\cA\subseteq\cL\text{ s.t. }|\cA|\leq [\alpha-1]^+. 
    \end{equation}
    This coincides with the sum-rate optimality condition (28) in~\cite{WS-SMDC-IT20}, which implies that superposition coding cannot be optimal for $2\leq s\leq L$. 
\end{remark}
\begin{remark}
    We can generalize the optimality of superposition to the following setup. Consider the $(L,s)$ sliding secure SMDC problem where $s$ divides $L$ and
    \begin{equation}
        H(X_j)=0,\text{ for }j\neq 0(\text{mod } s).
    \end{equation}
    That is we only have a set of $\frac{L}{s}$ sources $\bm{X}_{s},\bm{X}_{2s},\cdots,\bm{X}_{L}$. The sources are encoded by a set of $L$ encoders. For $1\leq \alpha\leq \frac{L}{s}$, the source $\bm{X}_{\alpha\cdot s}$ is required to be losslessly reconstructed by the decoder if any subset of $\alpha\cdot s$ encoders are accessible and should be kept perfectly secure from the eavesdropper if no more than $\alpha\cdot(s-1)$ encoders are accessible. Superposition remains to be optimal for this setup. The proof is very similar to that of \Cref{thm-L1}, we omit the proof here. This setup reduces to $(L,1)$ sliding secure SMDC when $s=1$.
\end{remark}


\section{Multilevel Secret Sharing}
\subsection{$(3,2)$ Multilevel Secret Sharing}  \label{section-32mss}
Consider the $(3,2)$ multilevel secret sharing problem. Now we have two independent sources $\bm{X}_2$ and $\bm{X}_3$. For $\alpha=2,3$, the source $\bm{X}_{\alpha}$ can be losslessly reconstructed by the decoder if a subset of $\alpha$ encoders are accessible and $\bm{X}_{\alpha}$ should be kept perfectly secure if the eavesdropper can access no more than $\alpha-2$ encoders. The reconstruction and security constraints in \eqref{recover constraint-mss}\eqref{security constraint-mss} become
\begin{align}
H(\bm{X}_2|W_i,W_j)&=0,\text{ for all } 1\leq i<j\leq 3  \label{recover constraint-mss32-2}\\
H(\bm{X}_3|W_1,W_2,W_3)&=0,  \label{recover constraint-mss32-3}\\
H(\bm{X}_3|W_i)&=H(\bm{X}_3),\text{ for all } i=1,2,3. \label{security constraint-mss32}
\end{align}
For $x,y\in\{1,2,3\}$, define the operation $\odot$ by
\begin{equation}
x\odot y=\begin{cases}x+y,&\text{ if }x+y\leq 3\\x+y-3&\text{ if }x+y>3.\end{cases}
\end{equation}
Let $\cR_1^*$ be the collection of nonnegative rate tuple $(R_1,R_2,R_3)$ such that 
\begin{align}
2R_i+R_j&\geq H(X_2)+H(X_3), \text{ for }1\leq i,j\leq 3 \text{ and }i\neq j \label{rate-mss32-1} \\
R_i+R_j&\geq \frac{1}{2}H(X_2)+H(X_3), \text{ for }1\leq i<j\leq 3  \label{rate-mss32-2-1} \\
R_i+R_j&\geq H(X_2), \text{ for }1\leq i<j\leq 3  \label{rate-mss32-2-2} \\
R_1+R_2+R_3&\geq \frac{3}{2}H(X_2)+H(X_3),  \label{rate-mss32-3} \\
2R_i+R_{i\odot 1}+R_{i\odot 2}&\geq 2H(X_2)+H(X_3), \text{ for }1\leq i\leq 3.   \label{rate-mss32-4}
\end{align}
For notational simplicity, we will use the following abbreviation in the sequel:
\begin{align}
m&=\max\left\{H(X_2),~\frac{1}{2}H(X_2)+H(X_3)\right\}.  \label{def-m1}
\end{align}
The following theorem fully characterizes the rate region $\cR_{3,2}^{\text{mss}}$ of the $(3,2)$ multilevel secret sharing problem.
\begin{theorem}\label{thm-32mss}
	$\cR_{3,2}^{\text{mss}}=\cR_1^*$. 
\end{theorem}

\begin{proof}
	The achievability and converse proofs are in Appendices \ref{proof-achievability-thm32mss} and \ref{proof-converse-thm32mss}.
\end{proof}
\Cref{thm-32mss} indicates that superposition is \textit{suboptimal} for the $(3,2)$ multilevel secret sharing problem. In order to achieve the optimality, we need to jointly encode the two sources. The main idea is that we can use $\bm{X}_2$ as the secret key for $\bm{X}_3$ to reduce coding rates. 

Consider the $(3,2)$ classical SMDC problem with $H(X_1)=0$. We can see that the constraints \eqref{rate-mss32-2-2}-\eqref{rate-mss32-4} exactly characterize the rate region of the classical SMDC problem. The rate region $\cR_{3,2}$ of the multilevel secret sharing problem can be obtained from the classical SMDC problem by adding two more constraints \eqref{rate-mss32-1} and \eqref{rate-mss32-2-1}. This is reasonable since the reconstruction requirements are the same for the two problems. The security constraints of multilevel secret sharing problem require the additional constraints \eqref{rate-mss32-1} and \eqref{rate-mss32-2-1}.

\subsection{The General $(L,s)$ Multilevel Secret Sharing}  \label{section-Ls-mss}
Throughout this subsection, let the first $s-1$ sources $\bm{X}_1,\bm{X}_2,\cdots,\bm{X}_{s-1}$ in the classical SMDC problem be constants. For any $(L,s)$, since the reconstruction requirements of the multilevel secret sharing and classical SMDC problems are the same, the rate constraints of classical SMDC must be also satisfied by the multilevel secret sharing problem. Thus the rate region of classical SMDC provides an outer bound for the multilevel secret sharing problem. We state the result in the following lemma.
\begin{lemma} \label{lemma-subset-mss}
	For any $(L,s)$, $\cR_{L,s}^{\text{mss}}\subseteq\cR_{\text{SMDC}}^{L*}$.
\end{lemma}
\begin{proof}
	The containment is trivial.
\end{proof}
Since there are additional security requirements for the multilevel secret sharing problem, the outer bound is not tight in general. However, from \Cref{section-32mss}, we can see that the minimum sum rates for the $(3,2)$ multilevel secret sharing problem and the classical SMDC problem are the same. It is natural to ask that whether the minimum sum rates for the general $(L,s)$ multilevel secret sharing problem and the classical SMDC problem are always the same. The answer is given in following theorem.
\begin{theorem} \label{thm-min_sum_rate-mss}
	For any $2\leq s\leq L$, if a rate tuple $(R_1,R_2,\cdots,R_L)$ is admissible for the $(L,s)$ multilevel secret sharing problem, then
	\begin{equation}
	R_1+R_2+\cdots+R_L\geq \sum_{\alpha=s}^{L}\frac{L}{\alpha}H(X_{\alpha}).  \label{min_sum_rate-mss}
	\end{equation}
	Moreover, this lower bound is tight.
\end{theorem}

\begin{proof}
	The converse part follows from \Cref{lemma-subset-mss}. 
	The details for proving the converse of the sum rate bound can be found in \cite{yeung97}.
	
	Next, we propose a coding scheme that achieves the minimum sum rate in \eqref{min_sum_rate-mss}. 
	Let $q$ be the smallest prime such that $q\geq L$. 
	For $\alpha\in\cL$, without loss of generality, we assume that $\bm{X}_{\alpha}$ is 
	a memoryless uniformly distributed sequence of length $l_{\alpha}$ over $\mathbb{F}_q$. 
	For $1\leq i\leq l_{\alpha}$, denote the $i$-th symbol of $\bm{X}_{\alpha}$ by $X_{\alpha,i}$. 
	For sufficiently large $n$, the length $l_{\alpha}$ is arbitrary close to $\frac{n H(X_{\alpha})}{\log_2 q}$. 
	Equipartition the sequence $\bm{X}_{\alpha}$ into $\alpha$ mutually independent pieces, 
	denoted by $\bm{X}_{\alpha}^1, \bm{X}_{\alpha}^2, \cdots, \bm{X}_{\alpha}^{\alpha}$. 
	Without loss of generality, assume 
	\begin{equation}
	\bm{X}_{\alpha}^{i}=\left(X_{\alpha,(i-1)\frac{l_{\alpha}}{\alpha}+1},X_{\alpha,(i-1)\frac{l_{\alpha}}{\alpha}+2},\cdots,X_{\alpha,i\cdot\frac{l_{\alpha}}{\alpha}}\right)
	\end{equation}
	for $1\leq i\leq \alpha$.
	
	Choose any $L$ distinct nonzero elements from $\mathbb{F}_q$, denoted by $\{b_1,b_2,\cdots,b_{L}\}$. Define a Vandermond matrix by 
	\begin{equation}
	V_{\alpha\times L}=\left[\begin{array}{cccc}1&1&\cdots&1\\ b_1&b_2&\cdots&b_{L}\\ b_1^2&b_2^2&\cdots&b_{L}^2\\ \vdots&\vdots&\ddots&\vdots\\ b_1^{\alpha-1}&b_2^{\alpha-1}&\cdots&b_{L}^{\alpha-1}\end{array}\right].
	\end{equation}
	Then any $\alpha$ columns of $V_{\alpha\times L}$ are linearly independent. Thus we can construct an $(L,\alpha)$ MDS code with the generator matrix $V_{\alpha\times L}$. Let
	\begin{equation}
	\left(\bm{Y}_{\alpha}^1,\bm{Y}_{\alpha}^2,\cdots,\bm{Y}_{\alpha}^L\right)=\left(\bm{X}_{\alpha}^1,\bm{X}_{\alpha}^2,\cdots,\bm{X}_{\alpha}^{\alpha}\right)\cdot V_{\alpha\times L}. \label{MDS-encode}
	\end{equation}
	For any $s\leq \alpha\leq L$ and $l\in\cL$, the length of $\bm{Y}_{\alpha}^l$ is $\frac{l_{\alpha}}{\alpha}$ and we have
	\begin{equation}
	H(\bm{Y}_{\alpha}^l)\leq H(\bm{X}_{\alpha}^l)=\frac{1}{\alpha}H(X_{\alpha}). \label{upper-bound_entropy-Y}
	\end{equation}
	For $1\leq i\leq \frac{l_{\alpha}}{\alpha}$, denote the $i$-th symbol of $\bm{Y}_{\alpha}^l$ by $Y_{\alpha,i}^{l}$. Since the operation on $\bm{X}_{\alpha}^i~(1\leq i\leq \alpha)$ is symbolwise, all the symbols of $\bm{Y}_{\alpha}^l$ are mutually independent and thus 
	\begin{equation}
	H(\bm{Y}_{\alpha}^l)=\sum_{i=1}^{l_{\alpha}/\alpha}H(Y_{\alpha,i}^{l}).
	\end{equation}
    The MDS property of \eqref{MDS-encode} ensures that any $\alpha$ of $\left\{\bm{Y}_{\alpha}^1,\bm{Y}_{\alpha}^2,\cdots,\bm{Y}_{\alpha}^L\right\}$ can losslessly recover the sequence $\bm{X}_{\alpha}^1,\bm{X}_{\alpha}^2,\cdots,\bm{X}_{\alpha}^\alpha$ and then can recover the source $\bm{X}_{\alpha}$. This implies that 
	\begin{equation}
	\sum_{l\in\cB}H(\bm{Y}_{\alpha}^{l})\geq H(\bm{X}_{\alpha}), ~\forall \cB\subseteq\cL, |\cB|=\alpha.   \label{lower-bound_entropy-Y}
	\end{equation}
	From \eqref{upper-bound_entropy-Y} and \eqref{lower-bound_entropy-Y}, we have
	\begin{equation}
	H(\bm{Y}_{\alpha}^{l})=\frac{1}{\alpha}H(\bm{X}_{\alpha}).
	\end{equation}
	Thus for $l\in\cL$ and $1\leq i\leq \frac{l_{\alpha}}{\alpha}$ 
	\begin{equation}
	H(Y_{\alpha,i}^{l})=\frac{1}{l_{\alpha}}H(\bm{X}_{\alpha})=\log_2q,
	\end{equation}
	which implies that $Y_{\alpha,i}^{l}$ is uniformly distributed.
	
	Equipartition $\bm{Y}_{\alpha}^{l}$ into $L(\alpha-s+1)$ mutually independent pieces, denoted by $\bm{Y}_{\alpha}^{l,1},\bm{Y}_{\alpha}^{l,2},\cdots, \bm{Y}_{\alpha}^{l,L(\alpha-s+1)}$. For $1\leq i\leq L(\alpha-s+1)$, let 
	\begin{equation}
	\bm{Y}_{\alpha}^{l,i}=\left(Y_{\alpha,\frac{(i-1)\cdot l_{\alpha}}{\alpha L(\alpha-s+1)}+1}^l,Y_{\alpha,\frac{(i-1)\cdot l_{\alpha}}{\alpha L(\alpha-s+1)}+2}^l,\cdots,Y_{\alpha,\frac{i\cdot l_{\alpha}}{\alpha L(\alpha-s+1)}}^l\right)
	\end{equation}
	for $1\leq i\leq L(\alpha-s+1)$.
	
	Now we construct the code that achieves the minimum sum rate. Since there is no ambiguity, we simply use $``+"$ to denote the addition of two sequences over $\mathbb{F}_q$ and $\oplus$ to denote the modulo $L$ addition. We first consider a special case that, for all $s+1\leq \alpha\leq L$,
	\begin{equation}
	\frac{1}{\alpha-1}H(X_{\alpha-1})=\frac{L(\alpha-s)}{\alpha}H(X_{\alpha}),
	\end{equation}
	which is equivalent to 
	\begin{equation}
	\frac{l_{\alpha-1}}{\alpha-1}=L(\alpha-s)\cdot\frac{l_{\alpha}}{\alpha}.
	\end{equation}
	The encoding and decoding procedures are as follows.
	\begin{itemize}
		\item Encoding: For $l\in\cL$, the output of Encoder-$l$ is
		\begin{align}
		W_l&=\left(\bm{Y}_{s}^{l},\bm{Y}_{s+1}^{l}+\bm{Y}_s^{l\oplus 1,l},\bm{Y}_{s+2}^{l}+\sum_{i=1}^2\bm{Y}_{s+1}^{l\oplus i,2(l-1)+i},\cdots,\right.  \nonumber \\
		&\hspace{0.5cm}\left.\bm{Y}_{\alpha}^l+\sum_{i=1}^{\alpha-s}\bm{Y}_{\alpha-1}^{l\oplus i,(l-1)(\alpha-s)+i}, \cdots,\bm{Y}_{L}^{l}+\sum_{i=1}^{L-s}\bm{Y}_{L-1}^{l\oplus i, (l-1)(L-s)+i}\right).  \label{coding_scheme-mss-Ls}
		\end{align}
		
		\item Decoding: When receiving a subset $\cU$ of codewords such that $|\cU|=\alpha$, we do the following:
		\begin{enumerate}[i.]
			\item If $\alpha=s$, the decoding of $\bm{X}_{\alpha}$ is the trivial MDS decoding.
			
			\item If $\alpha> s$, we first recover losslessly the sequence $\bm{X}_{s}$. Then initialize $j=s$.
			
			\item Calculate $\{\bm{Y}_j^1,\bm{Y}_j^2,\cdots,\bm{Y}_j^L\}$ using \eqref{MDS-encode}, then recover $\{\bm{Y}_{j+1}^{l}:l\in\cU\}$ from $\{\bm{Y}_{j+1}^{l}+\sum_{i=1}^{j-s+1}\bm{Y}_j^{l\oplus i,(l-1)(j-s+1)+i}:l\in\cU\}$. Since $\alpha\geq j+1$, the decoder can recover losslessly the sequence $\bm{X}_{j+1}$ and then the source $\bm{X}_{j+1}$.
			
			\item If $\alpha=j+1$, we are done. If $\alpha>j+1$, set $j=j+1$ and then go back to step (iii). 
		\end{enumerate}
		\textbf{Note:} For $s+1\leq \alpha\leq L$, decoding of the source $\bm{X}_{\alpha}$ should base on total information of all the previous sources $\bm{X}_s,\bm{X}_{s+1},\cdots,\bm{X}_{\alpha-1}$. 
	\end{itemize}
	
Next we check the security requirements in \eqref{security constraint-mss}. For $s+1\leq \alpha\leq L$, the eavesdropper can access a subset $\cA$ of encoders such that $|\cA|=\alpha-s$. Without loss of generality, assume $\cA=\{1,2,\cdots,\alpha-s\}$. The eavesdropper can recover at most $\bm{Y}_{\alpha-1}^1,\bm{Y}_{\alpha-1}^2,\cdots,\bm{Y}_{\alpha-1}^{\alpha-s}$ of the $(\alpha-1)$-th source. Any $\alpha-1$ of $\bm{Y}_{\alpha-1}^1,\bm{Y}_{\alpha-1}^2,\cdots,\bm{Y}_{\alpha-1}^{L}$ are mutually independent. Then we have
\begin{align}
&H\big(\sum_{i=1}^{\alpha-s}\bm{Y}_{\alpha-1}^{1\oplus i,i}, \sum_{i=1}^{\alpha-s}\bm{Y}_{\alpha-1}^{2\oplus i,(\alpha-s)+i}, \cdots,\sum_{i=1}^{\alpha-s}\bm{Y}_{\alpha-1}^{(\alpha-s)\oplus i,(\alpha-s-1)(\alpha-s)+i} |\bm{Y}_{\alpha-1}^{1},\bm{Y}_{\alpha-1}^{2},\cdots,\bm{Y}_{\alpha-1}^{\alpha-s}\big) \nonumber \\
&=\sum_{l=1}^{\alpha-s}H\big(\sum_{i=1}^{\alpha-s}\bm{Y}_{\alpha-1}^{l\oplus i,(l-1)(\alpha-s)+i}|\bm{Y}_{\alpha-1}^{1},\bm{Y}_{\alpha-1}^{2},\cdots,\bm{Y}_{\alpha-1}^{\alpha-s},\sum_{i=1}^{\alpha-s}\bm{Y}_{\alpha-1}^{1\oplus i,i},\cdots,\sum_{i=1}^{\alpha-s}\bm{Y}_{\alpha-1}^{(l-1)\oplus i,(l-2)(\alpha-s)+i}\big)  \nonumber \\
&=\sum_{l=1}^{\alpha-s}H\big(\bm{Y}_{\alpha-1}^{\alpha-s+1,l(\alpha-s-1)+1}+\bm{Y}_{\alpha-1}^{\alpha-s+2,l(\alpha-s-1)+2}+\cdots+ \bm{Y}_{\alpha-1}^{\min\{\alpha-s+l,L\},\min\{l(\alpha-s),(l-1)(\alpha-s)+L-l\}}|    \nonumber \\
&\quad \bm{Y}_{\alpha-1}^{1},\bm{Y}_{\alpha-1}^{2},\cdots,\bm{Y}_{\alpha-1}^{\alpha-s},\sum_{i=1}^{\alpha-s}\bm{Y}_{\alpha-1}^{1\oplus i,i},\cdots,\sum_{i=1}^{\alpha-s}\bm{Y}_{\alpha-1}^{(l-1)\oplus i,(l-2)(\alpha-s)+i}\big)    \label{independent-1-3} \\
&=\sum_{l=1}^{\alpha-s}H\big(\bm{Y}_{\alpha-1}^{\alpha-s+1,l(\alpha-s-1)+1}+\bm{Y}_{\alpha-1}^{\alpha-s+2,l(\alpha-s-1)+2}+\cdots+ \bm{Y}_{\alpha-1}^{\min\{\alpha-s+l,L\},\min\{l(\alpha-s),(l-1)(\alpha-s)+L-l\}}|    \nonumber \\
&\quad \bm{Y}_{\alpha-1}^{1,l(\alpha-s-1)+1},\bm{Y}_{\alpha-1}^{2,l(\alpha-s-1)+1},\cdots,\bm{Y}_{\alpha-1}^{\alpha-s,l(\alpha-s-1)+1},\bm{Y}_{\alpha-1}^{1,l(\alpha-s-1)+2},\bm{Y}_{\alpha-1}^{2,l(\alpha-s-1)+2},\cdots, \bm{Y}_{\alpha-1}^{\alpha-s,l(\alpha-s-1)+2}  \nonumber \\
&\quad \bm{Y}_{\alpha-1}^{1,\min\{l(\alpha-s),(l-1)(\alpha-s)+L-l\}},\bm{Y}_{\alpha-1}^{2,\min\{l(\alpha-s),(l-1)(\alpha-s)+L-l\}},\cdots,\bm{Y}_{\alpha-1}^{\alpha-s,\min\{l(\alpha-s),(l-1)(\alpha-s)+L-l\}}\big)    \label{independent-1-4} \\
&=\sum_{l=1}^{\alpha-s}H\big(\bm{Y}_{\alpha-1}^{\alpha-s+1,l(\alpha-s-1)+1}+\bm{Y}_{\alpha-1}^{\alpha-s+2,l(\alpha-s-1)+2}+\cdots+ \bm{Y}_{\alpha-1}^{\min\{\alpha-s+l,L\},\min\{l(\alpha-s),(l-1)(\alpha-s)+L-l\}}\big)
\label{independent-1-5} \\
&=\sum_{l=1}^{\alpha-s}H\big(\sum_{i=1}^{\alpha-s}\bm{Y}_s^{l\oplus i,(l-1)(\alpha-s)+i}\big)  \label{independent-1}
\end{align}
where \eqref{independent-1-3} follows from $H(A+B|A)=H(B|A)$, \eqref{independent-1-4} follows from the mutual independence of the partitions of $\bm{Y}_{\alpha-1}^i$ for any $i\in\cL$, \eqref{independent-1-5} follows from the mutual independence of any $\alpha-1$ of $\bm{Y}_{\alpha-1}^1,\bm{Y}_{\alpha-1}^2,\cdots,\bm{Y}_{\alpha-1}^{L}$ and \eqref{independent-1} follows from the fact that all symbols of $\bm{Y}_{\alpha-1}$ and the summations of any $j~(j\leq \alpha-1)$ of them are uniformly distributed. Thus, we have
\begin{align}
&H\big(\bm{Y}_{\alpha}^{1},\bm{Y}_{\alpha}^{2},\cdots,\bm{Y}_{\alpha}^{\alpha-s},\sum_{i=1}^{\alpha-s}\bm{Y}_{\alpha-1}^{1\oplus i,i},\cdots,\sum_{i=1}^{\alpha-s}\bm{Y}_{\alpha-1}^{(\alpha-s)\oplus i,(\alpha-s-1)(\alpha-s)+i},\bm{Y}_{\alpha-1}^{1},\bm{Y}_{\alpha-1}^{2},\cdots,\bm{Y}_{\alpha-1}^{\alpha-s}\big)   \nonumber \\ 
&=\sum_{l=1}^{\alpha-s}H(\bm{Y}_{\alpha-1}^l)+H\big(\bm{Y}_{\alpha}^{1},\bm{Y}_{\alpha}^{2},\cdots,\bm{Y}_{\alpha}^{\alpha-s}|\bm{Y}_{\alpha-1}^{1},\bm{Y}_{\alpha-1}^{2},\cdots,\bm{Y}_{\alpha-1}^{\alpha-s}\big)  \nonumber \\
&\quad +H\big(\sum_{i=1}^{\alpha-s}\bm{Y}_{\alpha-1}^{1\oplus i,i},\cdots,\sum_{i=1}^{\alpha-s}\bm{Y}_{\alpha-1}^{(\alpha-s)\oplus i,(\alpha-s-1)(\alpha-s)+i} |\bm{Y}_{\alpha-1}^{1},\bm{Y}_{\alpha-1}^{2},\cdots,\bm{Y}_{\alpha-1}^{\alpha-s}, \bm{Y}_{\alpha}^{1},\bm{Y}_{\alpha}^{2},\cdots,\bm{Y}_{\alpha}^{\alpha-s}\big)  \nonumber \\ 
&=\sum_{l=1}^{\alpha-s}H(\bm{Y}_{\alpha-1}^l)+H\big(\bm{Y}_{\alpha}^{1},\bm{Y}_{\alpha}^{2},\cdots,\bm{Y}_{\alpha}^{\alpha-s}\big)  \nonumber \\
&\quad +H\big(\sum_{i=1}^{\alpha-s}\bm{Y}_{\alpha-1}^{1\oplus i,i},\cdots,\sum_{i=1}^{\alpha-s}\bm{Y}_{\alpha-1}^{(\alpha-s)\oplus i,(\alpha-s-1)(\alpha-s)+i} |\bm{Y}_{\alpha-1}^{1},\bm{Y}_{\alpha-1}^{2},\cdots,\bm{Y}_{\alpha-1}^{\alpha-s}\big)  \label{independent-2-1} \\ 
&=\sum_{l=1}^{\alpha-s}H(\bm{Y}_{\alpha-1}^l)+\sum_{l=1}^{\alpha-s}H\big(\bm{Y}_{\alpha}^{l}\big)+\sum_{l=1}^{\alpha-s}H\big(\sum_{i=1}^{\alpha-s}\bm{Y}_{\alpha-1}^{l\oplus i,(l-1)(\alpha-s)+i}\big)  \label{independent-2}
\end{align}
where \eqref{independent-2-1} follows from the mutual independence of $\bm{Y}_{\alpha}$ and $\bm{Y}_{\alpha-1}$ and \eqref{independent-2} follows from \eqref{independent-1} and the mutual independence of $\bm{Y}_{\alpha}^1,\bm{Y}_{\alpha}^1,\cdots,\bm{Y}_{\alpha}^{\alpha}$. This implies that all of $\bm{Y}_{\alpha}^{1},\bm{Y}_{\alpha}^{2},\cdots,\bm{Y}_{\alpha}^{\alpha-s}$, $\sum_{i=1}^{\alpha-s}\bm{Y}_{\alpha-1}^{1\oplus i,i}$, $\sum_{i=1}^{\alpha-s}\bm{Y}_{\alpha-1}^{2\oplus i,(\alpha-s)+i}, \cdots$, $\sum_{i=1}^{\alpha-s}\bm{Y}_{\alpha-1}^{(\alpha-s)\oplus i,(\alpha-s-1)(\alpha-s)+i}$, $\bm{Y}_{\alpha-1}^{1},\bm{Y}_{\alpha-1}^{2},\cdots,\bm{Y}_{\alpha-1}^{\alpha-s}$ are mutually independent. From Appendix \ref{proof-indep-2}, we have
\begin{align}
&\hspace{-0.7cm}H\big(\bm{Y}_{\alpha}^1,\bm{Y}_{\alpha}^2,\cdots,\bm{Y}_{\alpha}^{\alpha-s}|\bm{Y}_{\alpha}^1+\sum_{i=1}^{\alpha-s}\bm{Y}_{\alpha-1}^{1\oplus i,i}, \cdots, \bm{Y}_{\alpha}^{\alpha-s}+\sum_{i=1}^{\alpha-s}\bm{Y}_{\alpha-1}^{(\alpha-s)\oplus i,(\alpha-s-1)(\alpha-s)+i}, \bm{Y}_{\alpha-1}^{1},\bm{Y}_{\alpha-1}^2,\cdots,\bm{Y}_{\alpha-1}^{\alpha-s}\big)   \nonumber \\
&=H\big(\bm{Y}_{\alpha}^1,\bm{Y}_{\alpha}^2,\cdots,\bm{Y}_{\alpha}^{\alpha-s}\big),
\end{align}
which implies that
\begin{equation}
H\big(\bm{Y}_{\alpha}^1,\bm{Y}_{\alpha}^2,\cdots,\bm{Y}_{\alpha}^{\alpha-s}|W_1,\cdots,W_{\alpha-s}\big) =H\big(\bm{Y}_{\alpha}^1,\bm{Y}_{\alpha}^2,\cdots,\bm{Y}_{\alpha}^{\alpha-s}\big),
\end{equation}
and thus
\begin{equation}
H\big(\bm{X}_{\alpha}|W_{\cA}\big) =H\big(\bm{X}_{\alpha}\big).
\end{equation}
for $s+1\leq \alpha\leq L$. Thus, all security requirements in \eqref{security constraint-mss} are satisfied.
	
From \eqref{coding_scheme-mss-Ls}, we obtain the coding rates 
\begin{equation}
R_i=\sum_{\alpha=s}^{L}\frac{1}{\alpha}H(X_{\alpha}),\text{ for }i\in\cL,  \label{rate-coding-mss}
\end{equation}
which implies that
\begin{equation}
\sum_{i=1}^{L}R_i=\sum_{\alpha=s}^{L}\frac{L}{\alpha}H(X_{\alpha}).  \label{sum_rate-coding-mss}
\end{equation}
Thus, the coding scheme achieves the sum rate bound in \eqref{min_sum_rate-mss} if
\begin{equation}
\frac{1}{\alpha-1}H(X_{\alpha-1})=\frac{L(\alpha-s)}{\alpha}H(X_{\alpha}),
\end{equation}
for all $s+1\leq\alpha\leq L$. For the case that
\begin{equation}
\frac{1}{\alpha-1}H(X_{\alpha-1})\geq \frac{L(\alpha-s)}{\alpha}H(X_{\alpha})
\end{equation}
for all $s+1\leq \alpha\leq L$, we modify the coding scheme in \eqref{coding_scheme-mss-Ls} as follows. Denote the first $\frac{l_{\alpha}}{\alpha}$ symbols of $\bm{Y}_{\alpha-1}^i~(i\in\cL)$ by $\bm{Y}_{\alpha-1}^{i*}$. Equipartition  $\bm{Y}_{\alpha-1}^{i*}$ into $L(\alpha-s)$ pieces, denoted by $\bm{Y}_{\alpha}^{i*,1}, \bm{Y}_{\alpha}^{i*,2}, \cdots, \bm{Y}_{\alpha}^{i*,L(\alpha-s)}$. Then we get the new codewords by replacing $\bm{Y}_{\alpha}^l+\sum_{i=1}^{\alpha-s}\bm{Y}_{\alpha-1}^{l\oplus i,(l-1)(\alpha-s)+i}$ in \eqref{coding_scheme-mss-Ls} with $\bm{Y}_{\alpha}^l+\sum_{i=1}^{\alpha-s}\bm{Y}_{\alpha-1}^{(l\oplus i)*,(l-1)(\alpha-s)+i}$ for all $l\in\cL$. It is easy to check that the decoding and security requirements are fulfilled and the same minimum sum rate is achieved.
	
Thus, we have designed a coding scheme that achieves the minimum sum rate in \eqref{min_sum_rate-mss} when
\begin{equation}
\frac{1}{\alpha-1}H(X_{\alpha-1})\geq \frac{L(\alpha-s)}{\alpha}H(X_{\alpha})  \label{sufficient-condition}
\end{equation}
for all $s+1\leq \alpha\leq L$. Therefore, we finish proving \Cref{thm-min_sum_rate-mss}.
\end{proof}
\begin{remark}
    The minimum sum rate in \eqref{min_sum_rate-mss} can be achieved only for $s\geq 2$. For $s=1$, superposition coding is proved to be optimal in \Cref{thm-L1} and thus no source can perform as secret keys for other sources.
\end{remark}

\begin{remark}
    In the proof, we derive a sufficient condition in \eqref{sufficient-condition} that the minimum sum rate can be achieved. 
    However, the condition is not necessary in general. 
    In the coding scheme, the sequence $\bm{Y}_{\alpha}^l$ is partitioned into many pieces to ensure perfect secrecy. 
    But on the other hand, too many partitions result in that the sufficient condition \eqref{sufficient-condition} is too tough. 
    The necessary and sufficient condition can be far from that. 
    The following lemma provides a much better sufficient condition for $L<2s$.
\end{remark}

\begin{lemma} \label{lemma-minimum_sum_rate-mss-opt}
	If $L<2s$, the minimum sum rate in \eqref{min_sum_rate-mss} can be achieved when 
	\begin{equation}
	\frac{H(X_s)}{s}\geq \frac{H(X_{s+1})}{s+1}\geq \cdots\geq \frac{H(X_L)}{L}.   \label{opt-condition}
	\end{equation}
\end{lemma}
\begin{proof}
	We first consider a special case that
	\begin{equation}
	\frac{H(X_s)}{s}=\frac{H(X_{s+1})}{s+1}=\cdots=\frac{H(X_L)}{L}.
	\end{equation}
	which is equivalent to
	\begin{equation}
	\frac{l_s}{s}=\frac{l_{s+1}}{s+1}=\cdots=\frac{l_L}{L}.
	\end{equation}
	The encoding and decoding procedures are as follows.
	\begin{itemize}
		\item Encoding: For $l\in\cL$, the output of Encoder-$l$ is
		\begin{equation}
		W_l=\left(\bm{Y}_{s}^{l},\bm{Y}_{s+1}^{l}+\bm{Y}_s^{l\oplus 1},\bm{Y}_{s+2}^{l}+\bm{Y}_{s+1}^{l\oplus 1},\cdots,\bm{Y}_{L}^{l}+\bm{Y}_{L-1}^{l\oplus 1}\right).  \label{coding_scheme-mss-Ls-opt}
		\end{equation}
		
		\item Decoding: For $s+1\leq \alpha\leq L$, decoding of the source $\bm{X}_{\alpha}$ should base on the previous source $\bm{X}_{\alpha-1}$. When receiving a subset $\cU$ of codewords such that $|\cU|=\alpha$, we have the following decoding procedure:
		\begin{enumerate}[i.]
			\item  If $\alpha=s$, the decoding of $\bm{X}_{\alpha}$ is the trivial MDS decoding.
			
			\item If $\alpha> s$, we first recover losslessly the sequence $\bm{X}_{s}$. Initialize $j=s$.
			
			\item Calculate $\{\bm{Y}_j^{l\oplus 1}:l\in\cU\}$ using \eqref{MDS-encode}, then recover $\{\bm{Y}_{j+1}^{l}:l\in\cU\}$ from $\{\bm{Y}_{j+1}^{l}+\bm{Y}_j^{l\oplus 1}:l\in\cU\}$. Since $\alpha\geq j+1$, the decoder can recover losslessly the sequence $\bm{X}_{j+1}$.
			
			\item If $\alpha=j+1$, we are done. If $\alpha>j+1$, set $j=j+1$ and then go back to step (iii). 
		\end{enumerate}
	\end{itemize}
	
	Next we check the security requirements in \eqref{security constraint-mss}. For $s+1\leq \alpha\leq L$, the eavesdropper can access a subset $\cA$ of encoders such that $|\cA|=\alpha-s$. Since $L<2s$, we have $\alpha-s<s$. Without loss of generality, assume $\cA=\{1,2,\cdots,\alpha-s\}$.
	
	The eavesdropper can recover $\bm{Y}_s^1,\bm{Y}_s^2,\cdots,\bm{Y}_s^{\alpha-s}$. Using \eqref{coding_scheme-mss-Ls-opt}, it can then recover $\bm{Y}_{s+1}^1, \bm{Y}_{s+1}^2, \cdots, \bm{Y}_{s+1}^{\alpha-s-1}$ from $\bm{Y}_{s+1}^1+\bm{Y}_s^2, \bm{Y}_{s+1}^2+\bm{Y}_s^3, \cdots, \bm{Y}_{s+1}^{\alpha-s-1}+\bm{Y}_s^{\alpha-s}$. Since $\bm{X}_s$ and $\bm{X}_{s+1}$ are mutually independent and $\alpha-s+1\leq s$, we have that $\bm{Y}_{s+1}^{\alpha-s},\bm{Y}_s^1,\bm{Y}_s^2,\cdots,\bm{Y}_s^{\alpha-s},\bm{Y}_s^{\alpha-s+1}$ are mutually independent. Thus, from Appendix \cref{proof-indep-2} we have
	\begin{equation}
	H\big(\bm{Y}_{s+1}^{\alpha-s}|\bm{Y}_{s+1}^{\alpha-s}+\bm{Y}_s^{\alpha-s+1},\bm{Y}_s^1,\bm{Y}_s^2,\cdots,\bm{Y}_s^{\alpha-s}\big)=H\big(\bm{Y}_{s+1}^{\alpha-s}\big). \label{independent-3}
	\end{equation}
	This implies that
	\begin{equation}
	H\big(\bm{Y}_{s+1}^{\alpha-s}|W_{\cA}\big)=H\big(\bm{Y}_{s+1}^{\alpha-s}\big).
	\end{equation}
	Similarly, for $s+1\leq j\leq \alpha$, the eavesdropper can recover $\bm{Y}_{j}^1, \bm{Y}_{j}^2, \cdots, \bm{Y}_{j}^{\alpha-j}$ using \eqref{coding_scheme-mss-Ls-opt}. Since $\alpha-s+1\leq j-1$, $\bm{Y}_{j-1}^1,\bm{Y}_{j-1}^2,\cdots, \bm{Y}_{j-1}^{\alpha-s+1}$ are mutually independent. From Appendix \ref{proof-indep-2} we have
	\begin{align}
	&\hspace{-0.7cm}H\big(\bm{Y}_j^{\alpha-j+1},\bm{Y}_j^{\alpha-j+2},\cdots,\bm{Y}_j^{\alpha-s}|\bm{Y}_j^{\alpha-j+1}+\bm{Y}_{j-1}^{\alpha-j+2},\bm{Y}_j^{\alpha-j+2}+\bm{Y}_{j-1}^{\alpha-j+3},\cdots,\bm{Y}_j^{\alpha-s}+\bm{Y}_{j-1}^{\alpha-s+1}, \bm{Y}_{j-1}^1,\bm{Y}_{j-1}^2,\cdots,\bm{Y}_{j-1}^{\alpha-j+1}\big)  \nonumber \\
	&=H\big(\bm{Y}_j^{\alpha-j+1},\bm{Y}_j^{\alpha-j+2},\cdots,\bm{Y}_j^{\alpha-s}\big),
	\end{align}
	which implies that
	\begin{equation}
	H\big(\bm{Y}_j^{\alpha-j+1},\bm{Y}_j^{\alpha-j+2},\cdots,\bm{Y}_j^{\alpha-s}|W_{\cA}\big)=H\big(\bm{Y}_j^{\alpha-j+1},\bm{Y}_j^{\alpha-j+2},\cdots,\bm{Y}_j^{\alpha-s}\big).
	\end{equation}
	Then for $j=\alpha$, the eavesdropper can recover nothing and we have
	\begin{equation}
	H\big(\bm{Y}_{\alpha}^{1},\bm{Y}_{\alpha}^{2},\cdots,\bm{Y}_{\alpha}^{\alpha-s}|W_{\cA}\big)=H\big(\bm{Y}_{\alpha}^{1},\bm{Y}_{\alpha}^{2},\cdots,\bm{Y}_{\alpha}^{\alpha-s}\big),
	\end{equation}
	which implies
	\begin{equation}
	H(\bm{X}_{\alpha}|W_{\cA})=H(\bm{X}_{\alpha}). \label{coding-analysis-mss-opt-secure-case1-last}
	\end{equation}
	Thus, all security requirements in \eqref{security constraint-mss} are satisfied.
	
	From \eqref{coding_scheme-mss-Ls-opt}, we obtain the coding rates 
	\begin{equation}
	R_i=\sum_{\alpha=s}^{L}\frac{1}{\alpha}H(X_{\alpha}),\text{ for }i\in\cL,  \label{rate-coding-mss-opt}
	\end{equation}
	which implies that
	\begin{equation}
	\sum_{i=1}^{L}R_i=\sum_{\alpha=s}^{L}\frac{L}{\alpha}H(X_{\alpha}).  \label{sum_rate-coding-mss-opt}
	\end{equation}
	Thus, the coding scheme achieves the sum rate bound in \eqref{min_sum_rate-mss} for the case that $\frac{H(X_{s})}{s}=\frac{H(X_{s+1})}{s+1}=\cdots=\frac{H(X_L)}{L}$. For the other cases, we modify the coding scheme in \eqref{coding_scheme-mss-Ls} as follows.
	\begin{enumerate}
		\item If $\frac{H(X_{\alpha-1})}{\alpha-1}\geq \frac{H(X_{\alpha})}{\alpha}$ for all $s+1\leq \alpha\leq L$, denote the first $\frac{l_{\alpha}}{\alpha}$ symbols of $\bm{Y}_{\alpha-1}^i~(i\in\cL)$ by $\bm{Y}_{\alpha-1}^{i*}$. Then we get the new codewords by replacing $\bm{Y}_{\alpha}^{l}+\bm{Y}_{\alpha-1}^{l\oplus 1}$ in \eqref{coding_scheme-mss-Ls-opt} with $\bm{Y}_{\alpha}^l+\bm{Y}_{\alpha-1}^{(l\oplus 1)*}$. It is easy to check that the decoding and security requirements are fulfilled and the same minimum sum rate is achieved.
		
		\item If $\frac{H(X_{\alpha-1})}{\alpha-1}<\frac{H(X_{\alpha})}{\alpha}$ for some $s+1\leq \alpha\leq L$, the source $\bm{X}_{\alpha-1}$ is not sufficient to serve as the secret key for the whole source $\bm{X}_{\alpha}$ and the minimum sum rate cannot be achieved by our coding scheme. The following example illustrates why the sum rate exceeds the bound in \eqref{min_sum_rate-mss}.
		\begin{example}
			Partition $\bm{X}_{\alpha}$ into $\alpha+1$ pieces denoted by $\bm{X}_{\alpha}^{0*}, \bm{X}_{\alpha}^{1*}, \bm{X}_{\alpha}^{2*}, \cdots, \bm{X}_{\alpha}^{(\alpha)*}$. The length of $\bm{X}_{\alpha}^{0*}$ is $l_{\alpha}^0=l_{\alpha}-\frac{\alpha}{\alpha-1}l_{\alpha-1}$ and the length of $\bm{X}_{\alpha}^{i*}~(1\leq i\leq \alpha)$ is $l_{\alpha}^*=\frac{l_{\alpha-1}}{\alpha-1}$. Let
			\begin{equation}
			\left(\bm{Y}_{\alpha}^{1*}, \bm{Y}_{\alpha}^{2*},\cdots, \bm{Y}_{\alpha}^{L*}\right) =\left(\bm{X}_{\alpha}^{1*}, \bm{X}_{\alpha}^{2*}, \cdots, \bm{X}_{\alpha}^{(\alpha)*}\right)\cdot V_{\alpha\times L}. \label{MDS-encode*}
			\end{equation}
			We use a $(\alpha,s,L)$ ramp secret sharing code in \cite{yamamoto85} to encode $\bm{X}_{\alpha}^{0*}$ with random keys. Denote the output of the secret sharing code by $\bm{Y}_{\alpha}^{0,1},\bm{Y}_{\alpha}^{0,2},\cdots,\bm{Y}_{\alpha}^{0,L}$. For $i\in\cL$, replace $\bm{Y}_{\alpha}^{l}+\bm{Y}_{\alpha-1}^{l\oplus 1}$ in \eqref{coding_scheme-mss-Ls-opt} with $\left(\bm{Y}_{\alpha}^{l*}+\bm{Y}_{\alpha-1}^{l\oplus 1}, \bm{Y}_{\alpha}^{0,i}\right)$. For $i\in\cL$, the coding rate $R_i^{\alpha}$ for source $\bm{X}_{\alpha}$ is
			\begin{equation}
			R_i^{\alpha}=\frac{1}{\alpha-1}H(X_{\alpha-1})+\frac{1}{s}\left[H(X_{\alpha})-\frac{\alpha}{\alpha-1}H(X_{\alpha-1})\right].
			\end{equation}
			Since $\frac{H(X_{\alpha-1})}{\alpha-1}<\frac{H(X_{\alpha})}{\alpha}$, we have
			\begin{equation}
			\sum_{i=1}^LR_i^{\alpha}=\frac{L}{\alpha-1}H(X_{\alpha-1})+\frac{L}{s}\left[H(X_{\alpha})-\frac{\alpha}{\alpha-1}H(X_{\alpha-1})\right]>\frac{L}{\alpha}H(X_{\alpha}).
			\end{equation}
			Thus, the sum rate exceeds the bound in \eqref{min_sum_rate-mss}.
		\end{example}
		
	\end{enumerate}
	We have designed a coding scheme that achieves the minimum sum rate in \eqref{min_sum_rate-mss} for the case that
	\begin{equation}
	\frac{H(X_s)}{s}\geq \frac{H(X_{s+1})}{s+1}\geq \cdots\geq \frac{H(X_L)}{L}.
	\end{equation}
	Therefore, \Cref{lemma-minimum_sum_rate-mss-opt} is proved.
\end{proof}

For $L<2s$, \Cref{lemma-minimum_sum_rate-mss-opt} provides a sufficient condition that achieves the minimum sum rate bound. We conjecture that the condition in \eqref{opt-condition} is also necessary.
\begin{conjecture}
	For $L<2s$, the sum rate bound in \eqref{min_sum_rate-mss} is inactive if 
	\begin{equation}
	\frac{H(X_{\alpha-1})}{\alpha-1}<\frac{H(X_{\alpha})}{\alpha}
	\end{equation}
	for some $s+1\leq \alpha\leq L$.
\end{conjecture}
We will try to prove the conjecture in the future work. The rate region in \Cref{thm-32mss} provides an example that the conjecture is true, which is as follows. 
\begin{example}
	For $(L,s)=(3,2)$, the sum rate bound for the multilevel secret sharing problem in \eqref{rate-mss32-3} is inactive if and only if 
	\begin{equation}
	H(X_2)<\frac{2}{3}H(X_3).
	\end{equation}
	We can find the details in Appendix \ref{proof-achievability-thm32mss}.
\end{example}
Next, we show the \textit{suboptimality} of superposition coding for the $(L,s)$ multilevel secret sharing problem. For each $s\leq \alpha\leq L$, the problem of encoding the single source $\bm{X}_{\alpha}$ is the $(\alpha-s,\alpha,L)$ ramp secret sharing problem whose rate region is characterized by $\cR\big(L,s,H(X_{\alpha})\big)$ in \eqref{region-rss}. The superposition region $\cR_{\sup}^{2}$ induced by separately encoding the $L-s+1$ independent sources is the set of nonnegative rate tuples $(R_1,R_2,\cdots,R_L)$ such that
\begin{equation}
R_i=\sum_{\alpha=s}^{L}r_i^{\alpha}, \text{ for } i\in\cL
\end{equation}
where $r_i^{\alpha}\geq 0$, $s\leq \alpha\leq L$ and
\begin{equation}
\big(r_1^{\alpha},r_2^{\alpha},\cdots,r_L^{\alpha}\big)\in\cR\big(L,s,H(X_{\alpha})\big).
\end{equation}
It is easy to check that
\begin{equation}
\cR_{\sup}^{2}=\cR\big(L,s,\sum_{\alpha=s}^{L}H(X_{\alpha})\big).  \label{sup_region-mss}
\end{equation}

Then from \Cref{thm-min_sum_rate-mss}, we have the following corollary.
\begin{corollary} \label{cor-proper_subset-mss}
	$\cR_{\text{sup}}^2\subsetneq\cR_{L,s}^{\text{mss}}$ for $2\leq s\leq L-1$.
\end{corollary}

\begin{proof}
	From \eqref{sup_region-mss}, we can write the sum rate bound of the superposition region as follows,
	\begin{equation}
	R_1+R_2+\cdots+R_L\geq \sum_{\alpha=s}^{L}\frac{L}{s}H(X_{\alpha}).  \label{sup_min_sum_rate-mss}
	\end{equation}
	For $2\leq s\leq L-1$, since
	\begin{equation}
	\sum_{\alpha=s}^{L}\frac{L}{s}H(X_{\alpha})>\sum_{\alpha=s}^{L}\frac{L}{\alpha}H(X_{\alpha}),
	\end{equation}
	we conclude that superposition coding for $(L,s)$ multilevel secret sharing problem is suboptimal. Thus, \Cref{cor-proper_subset-mss} is proved.
\end{proof}

\begin{remark}
    For $s=1$, superposition is proved to be optimal in \Cref{section-L1}. 
    For $s=L$, the problem reduces to the classical SMDC problem with $H(X_1)=\cdots=H(X_{\alpha-1})=0$ and thus superposition is optimal.
\end{remark}

\section{Sliding Secure SMDC}
\subsection{$(3,2)$ Sliding Secure SMDC} \label{section-32sMDC}
Consider the $(3,2)$ sliding secure SMDC problem. Now we have three independent sources $\bm{X}_1$, $\bm{X}_2$, and $\bm{X}_3$. For $\alpha=1,2,3$, the source $\bm{X}_{\alpha}$ can be losslessly reconstructed by the decoder if a subset of $\alpha$ encoders are accessible and $\bm{X}_{\alpha}$ should be kept perfectly secure if the eavesdropper can access no more than $[\alpha-2]^+$ encoders. The reconstruction and security constraints in \eqref{recover constraint}\eqref{security constraint} become
\begin{align}
H(\bm{X}_1|W_i)&=0,\text{ for all } i=1,2,3  \label{recover constraint-3,2-1}\\
H(\bm{X}_2|W_i,W_j)&=0,\text{ for all } 1\leq i<j\leq 3  \label{recover constraint-3,2-2}\\
H(\bm{X}_3|W_1,W_2,W_3)&=0,  \label{recover constraint-3,2-3}\\
H(\bm{X}_3|W_i)&=H(\bm{X}_3),\text{ for all } i=1,2,3. \label{security constraint-3,2}
\end{align}

Let $\cR_2^*$ be the collection of nonnegative rate triples $(R_1,R_2,R_3)$ such that for $i=1,2,3$
\begin{equation}
R_i=r_i^0+r_i^1
\end{equation}
where $r_i^0, r_i^1>0$ and
\begin{align}
r_i^1&\geq H(X_1) \\
(r_1^0,r_2^0,r_3^0)&\in\cR_{3,2}^{\text{mss}}.
\end{align}
We can see that $\cR_2^*$ is the superposition region induced by separately encoding two sets of sources $\bm{X}_1$ and $(\bm{X}_2, \bm{X}_3)$ with rates $r_i^{1}$ and $r_i^{0}$, respectively. It is easy to write the region equivalently by eliminating $r_i^{j}$, which is the set of rate tuples $(R_1,R_2,R_3)$ such that
\begin{align}
R_i&\geq H(X_1), \text{ for }1\leq i\leq 3   \label{rate-sMDC32-1} \\
2R_i+R_j&\geq 3H(X_1)+H(X_2)+H(X_3), \text{ for }1\leq i,j\leq 3 \text{ and }i\neq j  \label{rate-sMDC32-2} \\
R_i+R_j&\geq 2H(X_1)+\frac{1}{2}H(X_2)+H(X_3), \text{ for }1\leq i<j\leq 3   \label{rate-sMDC32-3-1} \\
R_i+R_j&\geq 2H(X_1)+H(X_2), \text{ for }1\leq i<j\leq 3   \label{rate-sMDC32-3-2} \\
R_1+R_2+R_3&\geq 3H(X_1)+\frac{3}{2}H(X_2)+H(X_3),    \label{rate-sMDC32-4}\\
2R_i+R_{i\odot 1}+R_{i\odot 2}&\geq 4H(X_1)+2H(X_2)+H(X_3),  \text{ for }1\leq i\leq 3.   \label{rate-sMDC32-5}
\end{align}

The following theorem characterizes the rate region $\cR_{3,2}$ of the $(3,2)$ sliding secure SMDC problem. 
\begin{theorem}
	$\cR_{3,2}=\cR_2^*$. \label{thm-32-sMDC}
\end{theorem}
\begin{proof}
	Since $\cR_2^*$ is the rate region induced by superposition coding of $\bm{X}_1$ and $(\bm{X}_2,\bm{X}_3)$, all the rates $(R_1,R_2,R_3)\in\cR_2^*$ are achievable. We only need to show the converse. The details are in Appendix \ref{proof-thm-32-sMDC}.
\end{proof}
\Cref{thm-32-sMDC} indicates that separately encoding two sets of sources $\bm{X}_1$ and $(\bm{X}_2,\bm{X}_3)$ is optimal for the $(3,2)$ sliding secure SMDC problem even though superposition of three individual sources is suboptimal. The result coincides with the region of $(3,2)$ classical SMDC by setting $\bm{X}_3$ to be constants for both problems.

We can see that the constraints in \eqref{rate-sMDC32-1} and \eqref{rate-sMDC32-3-2}-\eqref{rate-sMDC32-5} are the same constraints that defines the classical SMDC region. Thus, the rate region $\cR_{3,2}$ can be obtained from the classical SMDC region by adding two more constraints \eqref{rate-sMDC32-2} and  \eqref{rate-sMDC32-3-1}.

\subsection{The General $(L,s)$ Sliding Secure SMDC}  \label{section-Ls-sMDC}
For any $(L,s)$, since the reconstruction requirements of the sliding secure SMDC and classical SMDC problems are the same, the rate constraints of classical SMDC must be also satisfied by the sliding secure SMDC problem. Thus the rate region of classical SMDC provides an outer bound for the sliding secure SMDC problem. We state the result in the following lemma.
\begin{lemma} \label{lemma-subset-sMDC}
	For any $(L,s)$, $\cR_{L,s}\subseteq\cR_{\text{SMDC}}^L$.
\end{lemma}
\begin{proof}
	The containment is trivial.
\end{proof}
Since there are additional security requirements for the sliding secure SMDC problem, the outer bound is not tight in general. However, from \Cref{section-32sMDC}, we can see that the $(3,2)$ sliding secure SMDC problem achieves the same minimum sum rate as the classical SMDC problem. Moreover, in \Cref{section-Ls-mss}, we show that the multilevel secret sharing problem achieves the sum rate bound of $\cR_{\text{SMDC}}^*$. In the following theorem, we generalize the sum rate property in \Cref{section-32sMDC,section-Ls-mss} to the general $(L,s)$ sliding secure SMDC problem.
\begin{theorem} \label{thm-min_sum_rate-sMDC}
	For any $2\leq s\leq L-1$, if a rate tuple $(R_1,R_2,\cdots,R_L)$ is admissible for the $(L,s)$ sliding secure SMDC problem, then
	\begin{equation}
	R_1+R_2+\cdots+R_L\geq \sum_{\alpha=1}^{L}\frac{L}{\alpha}H(X_{\alpha}).  \label{min_sum_rate-sMDC}
	\end{equation}
	Moreover, the lower bound is tight and can be achieved by superposition coding of the sets of sources $\bm{X}_1$, $\bm{X}_2$, $\cdots$, $\bm{X}_{s-1}$, and $(\bm{X}_s, \bm{X}_{s+1}, \cdots, \bm{X}_L)$.
\end{theorem}
\begin{proof}
	The converse follows from \Cref{lemma-subset-sMDC}.
	
	Next, we design a coding scheme that achieves that minimum sum rate in \eqref{min_sum_rate-sMDC}. For $i\in\cL$, and $0\leq \alpha\leq s-1$, let
	\begin{align}
	r_i^{\alpha}&=\frac{1}{\alpha}H(X_{\alpha}), \text{ for } 1\leq \alpha\leq s-1  \label{sup_rate-sMDC-1}\\
	r_i^0&=\sum_{\alpha=s}^{L}\frac{1}{\alpha}H(X_{\alpha}).  \label{sup_rate-sMDC-0}
	\end{align}
	Then separately encode the $s$ sets of sources $\bm{X}_1$, $\bm{X}_2$, $\cdots$, $\bm{X}_{s-1}$, and $(\bm{X}_s, \bm{X}_{s+1}, \cdots, \bm{X}_L)$ with rates $r_i^1,r_i^2,\cdots,r_i^{s-1}$, and $r_i^0$, respectively for $i\in\cL$. For any $1\leq \alpha\leq s-1$ and $\cB\subseteq\cL$ such that $|\cB|=\alpha$, since
	\begin{equation}
	\sum_{i\in\cB}r_i^{\alpha}=\sum_{i\in\cB}\frac{1}{\alpha}H(X_{\alpha})=H(X_{\alpha}),
	\end{equation}
	we can losslessly reconstruct the sources $\bm{X}_1,\bm{X}_2,\cdots,\bm{X}_{s-1}$. Since the rates $r_i^0~(i\in\cL)$ in \eqref{sup_rate-sMDC-0} are the same as rates in \eqref{rate-coding-mss}, we can use the coding scheme in \Cref{section-Ls-mss} to encode the set of sources $(\bm{X}_s,\bm{X}_{s+1},\cdots,\bm{X}_L)$ if
	\begin{equation}
	\frac{1}{\alpha-1}H(X_{\alpha-1})\geq \frac{L(\alpha-s)}{\alpha}H(X_{\alpha})
	\end{equation}
	for all $s+1\leq\alpha\leq L$. Thus, the rate tuple $(R_1,R_2,\cdots,R_L)$, where
	\begin{equation}
	R_i=\sum_{\alpha=0}^{s-1}r_i^{\alpha}, \text{ for }i\in\cL
	\end{equation}
	is admissible and superposition of the $s$ sets of sources achieves the sum rate 
	\begin{equation}
	R_1+R_2+\cdots+R_L=\sum_{i=1}^{L}\left(\sum_{\alpha=0}^{s-1}r_i^{\alpha}\right)=\sum_{i=1}^L\sum_{\alpha=1}^{L}\frac{1}{\alpha}H(X_{\alpha})=\sum_{\alpha=1}^{L}\frac{L}{\alpha}H(X_{\alpha}).
	\end{equation}
	Therefore, the optimality of such superposition coding in terms of achieving the minimum sum rate in \eqref{min_sum_rate-sMDC} is proved. This completes the proof of \Cref{thm-min_sum_rate-sMDC}.
\end{proof}

Let $\cR_{\sup}^{3}$ be the superposition region of the $(L,s)$ sliding secure SMDC problem induced by separately encoding the $L$ independent sources. Then $\cR_{\sup}^{3}$ is the set of rate tuples $(R_1,R_2,\cdots,R_L)$ such that for $i\in\cL$
\begin{equation}
R_i=\sum_{\alpha=1}^{L}r_i^{\alpha}
\end{equation}
where $r_i^{\alpha}\geq 0$, $1\leq \alpha\leq L$ and
\begin{align}
\sum_{i\in\cB}r_{i}^{\alpha}&\geq H(X_{\alpha}), \text{ for }1\leq \alpha\leq s-1 \text{ and }\cB\subseteq\cL \text{ s.t. }|\cB|=\alpha \\
\sum_{i\in\cB}r_{i}^{\alpha}&\geq H(X_{\alpha}), \text{ for }s\leq \alpha\leq L \text{ and }\cB\subseteq\cL \text{ s.t. }|\cB|=s.
\end{align}
It is easy to characterize the sum rate bound of $\cR_{\sup}^{3}$ by
\begin{align}
R_1+R_2+\cdots+R_L\geq \sum_{\alpha=1}^{s-1}\frac{L}{\alpha}H(X_{\alpha})+\sum_{\alpha=s}^{L}\frac{L}{s}H(X_{\alpha}).  \label{sup_region-sMDC}
\end{align}

\begin{corollary} \label{cor-proper_subset-sMDC}
	$\cR_{\text{sup}}^{3}\subsetneq\cR_{L,s}$ for $2\leq s\leq L-1$.
\end{corollary}

\begin{proof}
	The sum rate bound of $\cR_{\sup}^{3}$ is in \eqref{sup_region-sMDC}. For $2\leq s\leq L-1$, since
	\begin{equation}
	\sum_{\alpha=1}^{s-1}\frac{L}{\alpha}H(X_{\alpha})+\sum_{\alpha=s}^{L}\frac{L}{s}H(X_{\alpha})>\sum_{\alpha=1}^{L}\frac{L}{\alpha}H(X_{\alpha}),
	\end{equation}
	we conclude that superposition coding for $(L,s)$ sliding secure SMDC problem is suboptimal. This proves \Cref{cor-proper_subset-sMDC}.
\end{proof}
\begin{remark}
    For $s=1$, superposition is proved to be optimal in \Cref{section-L1}. 
    For $s=L$, the problem reduces to the classical SMDC problem and thus superposition of the $L$ sources is optimal.
\end{remark}

Even though \Cref{cor-proper_subset-sMDC} states that superposition of $\bm{X}_1$, $\bm{X}_2$, $\cdots$, $\bm{X}_L$ is suboptimal, \Cref{thm-min_sum_rate-sMDC} tells us that superposition of $\bm{X}_1$, $\bm{X}_2$, $\cdots$, $\bm{X}_{s-1}$, $(\bm{X}_s,\bm{X}_{s+1},\cdots,\bm{X}_L)$ achieves the minimum sum rate. Denote the superposition region induced by separately encoding the $s$ set of sources by $\cR_{\sup}^{4}$. Then $\cR_{\sup}^{4}$ is the set of nonnegative rate tuples $(R_1,R_2,\cdots,R_L)$ such that
\begin{equation}
R_i=\sum_{\alpha=0}^{s-1}r_i^{\alpha}, \text{ for } i\in\cL
\end{equation}
where $r_i^{\alpha}\geq 0$, $0\leq \alpha\leq s-1$ and
\begin{align}
\sum_{i\in\cB}r_{i}^{\alpha}&\geq H(X_{\alpha}), \text{ for }1\leq \alpha\leq s-1 \text{ and }\cB\subseteq\cL \text{ s.t. }|\cB|=\alpha \\
\big(r_1^0,r_2^0,\cdots,r_L^0\big)&\in\cR_{L,s}^{\text{mss}}.
\end{align}
We conjecture that superposition of $\bm{X}_1$, $\bm{X}_2$, $\cdots$, $\bm{X}_{s-1}$, $(\bm{X}_s,\bm{X}_{s+1},\cdots,\bm{X}_L)$ can achieve the entire admissible rate region of the sliding secure SMDC problem.
\begin{conjecture}
	$\cR_{L,s}=\cR_{\sup}^{4}$ for all $(L,s)$.
\end{conjecture}
A trivial case that the conjecture is true is the classical SMDC problem with $H(X_{s+1})=\cdots=H(X_L)=0$. The simplest nontrivial example is the $(3,2)$ problem whose rate region is characterized in \Cref{thm-32-sMDC}. The conjecture is also true for the special cases that $s=1$ and $s=L$.

However, since superposition is suboptimal for general $(L,s)$ multilevel secret sharing problem (with $2\leq s\leq L-1$), it is difficult to characterize the rate region $\cR_{L,s}^{\text{mss}}$. Thus, it is challenging to prove the conjecture at even the first step. Similar as the rate region of classical SMDC problem in \cite{yeung99}, we may try to find some implicit ways to characterize the rate region of sliding secure SMDC in the future work.

\section{Conclusion}  \label{section-conclusion}
This paper considered the problem of sliding secure SMDC, which is a generalization of the classical SMDC problem to the security settings. The $(L,s)$ sliding secure SMDC problem is specialized to the $(L,s)$ multilevel secret sharing problem when the first $s-1$ sources are constants. We have fully characterized the rate regions of the sliding secure SMDC and multilevel secret sharing problems for $s=1$ and $(L,s)=(3,2)$. For $s=1$, separately encoding independent sources (superposition coding) is optimal for both sliding secure SMDC and multilevel secret sharing problems.

For $s\geq 2$, it was shown that superposition coding of the independent sources is suboptimal for the multilevel secret sharing problem. The main idea that joint encoding can reduce coding rates is that we can use the previous source $\bm{X}_{\alpha-1}$ as the secret keys for source $\bm{X}_{\alpha}$. Based on this idea, we designed a coding scheme that achieves the entire rate region of the $(3,2)$ problem. For the general case, we proposed a coding scheme that achieves the minimum sum rate. However, it is difficult to determine whether such coding schemes can achieve the entire rate region since it is really challenging to characterize the region of the general multilevel secret sharing problem.

For the $(3,2)$ sliding secure SMDC problem, superposition of two sets of sources $\bm{X}_1$ and $(\bm{X}_2,\bm{X}_3)$ was shown to be optimal. For the general problem that $s\geq 2$, we have shown that superposition of the $s$ sets of sources $\bm{X}_1$, $\bm{X}_2$, $\cdots$, $\bm{X}_{s-1}$ and $(\bm{X}_s,\bm{X}_{s+1},\cdots,\bm{X}_{L})$ achieves the minimum sum rate. To show the optimality of achieving the entire rate region, we need more efforts in the future work.

\begin{appendices}
	\section{Achievability Proof of \Cref{thm-32mss}} \label{proof-achievability-thm32mss}
	In this section, we will show the achievability of $\cR_1^*$ by designing coding schemes for the following four case, respectively.
	\begin{enumerate}[i.]
		\item $m=\frac{1}{2}H(X_2)+H(X_3)$, $\frac{3}{2}m> \frac{3}{2}H(X_2)+H(X_3)$, and $2m> 2H(X_2)+H(X_3)$,
		\item $m=\frac{1}{2}H(X_2)+H(X_3)$, $\frac{3}{2}m\leq\frac{3}{2}H(X_2)+H(X_3)$, and $2m> 2H(X_2)+H(X_3)$,
		\item $m=\frac{1}{2}H(X_2)+H(X_3)$, $\frac{3}{2}m\leq\frac{3}{2}H(X_2)+H(X_3)$, and $2m\leq 2H(X_2)+H(X_3)$,
		\item $m=H(X_2)$.
	\end{enumerate}
	For each case, the rate region $\cR_1^*$ is a convex polyhedron specified by several hyperplanes. 
	Due to time-sharing arguments, we only need to show the achievability for the corner points.
	For $\alpha\in\cL$, without loss of generality, we assume that $\bm{X}_{\alpha}$ is 
	a memoryless binary symmetric sequence of length $l_{\alpha}$. 
	For sufficiently large $n$, $l_{\alpha}$ is arbitrary close to $n H(X_{\alpha})$. 
	
	\subsection{Case i} \label{proof-achievability-thm32mss-case1}
\begin{figure}[!h]
		\centering
		\begin{tikzpicture}[font=\scriptsize, scale=1.0]
		\draw [->,opacity=0.2] (0,0)--(-3.0,-3.0) node [left,black,opacity=.5](R1) {$R_1$};
		\draw [->,opacity=0.2] (0,0)--(6.5,0) node [right,black,opacity=.5](R2) {$R_2$};
		\draw [->,opacity=0.2] (0,0)--(0,5.0) node [right,black,opacity=.5](R3) {$R_3$};
		\fill [black,opacity=.5] (0,0) circle (1.0pt);
		
		\coordinate (O) at (0.2,0.1);
		\coordinate (Q1) at (2.8,2.5);
		\coordinate (Q2) at (1.5,-2.2);
		\coordinate (Q3) at (-2.3,0.7);
		\coordinate (P1) at (1.8,1.3);
		\coordinate (P2) at (1.0,-1.0);
		\coordinate (P3) at (-1.0,0.5);
		\coordinate (S1) at (1.0,0.8);
		\coordinate (S2) at (0.7,-0.6);
		\coordinate (S3) at (-0.4,0.3);
		
		\coordinate (Q1') at (2.8,4.5);
		\coordinate (Q1'') at (5.0,2.5);
		\coordinate (Q2') at (4.5,-2.2);
		\coordinate (Q2'') at (-0.5,-4.2);
		\coordinate (Q3') at (-4.6,-1.6);
		\coordinate (Q3'') at (-2.3,3.5);
		
		\coordinate (P1') at (4.5,1.3);
		\coordinate (P1'') at (1.8,3.5);
		\coordinate (P2') at (4.0,-1.0);
		\coordinate (P2'') at (-1.0,-3.0);
		\coordinate (P3') at (-1.0,3.5);
		\coordinate (P3'') at (-3.0,-1.5);
		
		\draw (O)--(P1) (O)--(P2) (O)--(P3);
		\foreach \j in {1,2,3}{\draw (Q\j)--(Q\j') (Q\j)--(Q\j'') (P\j)--(P\j') (P\j)--(P\j'') (Q\j)--(P\j);}
		
		\foreach \point in {Q2,Q3,P2,P3}{\fill [black,opacity=.5] (\point) circle (1.5pt);}
		\foreach \point in {O,Q1,P1}{\fill [red,opacity=.5] (\point) circle (1.5pt);}
		
		\node [below right](O) at (0.35,0) {$O=(\frac{1}{4}H_2+\frac{1}{2}H_3,\frac{1}{4}H_2+\frac{1}{2}H_3,\frac{1}{4}H_2+\frac{1}{2}H_3)$};
		\node [above right](Q1) at (Q1) {$Q_1=(0,H_2+H_3,H_2+H_3)$};
		\node [below right](Q2) at (Q2) {$Q_2=(H_2+H_3,H_2+H_3,0)$};
		\node [left](Q3) at (Q3) {$Q_3$};
		\node [below right](P1) at (P1) {$P_1=(\frac{1}{2}H_2,H_3,H_3)$};
		\node [below right](P2) at (P2) {$P_2=(H_3,H_3,\frac{1}{2}H_2)$};
		\node [below](P3) at (P3)  {$P_3$};
		
		\node [opacity=.5, right](R1) at (3.2,3.3) {$R_1\geq 0$};
		\node [opacity=.5, right](R3) at (1.5,-3.3) {$R_3\geq 0$};
		\node [opacity=.5, left](R2) at (-3.0,1.5) {$R_2\geq 0$};
		
		\node [opacity=.5, right](2R1R3) at (2.5,1.8) {$2R_1+R_3\geq H_2+H_3$};
		\node [opacity=.5, right](2R3R1) at (2.0,-1.8) {$R_1+2R_3\geq H_2+H_3$};
		\draw [opacity=0] (2.2,1.7)--(2.2,5.5) node [opacity=0.5,sloped,midway] {$2R_1+R_2\geq H_2+H_3$};
		\draw [opacity=0] (-1.6,1.0)--(-1.6,3.5) node [opacity=0.5,sloped,midway] {$R_1+2R_2\geq H_2+H_3$};
		\draw [opacity=0] (-3.7,-1.7)--(-1.5,0.5) node [opacity=0.5,sloped,midway] {$2R_2+R_3\geq H_2+H_3$};
		\draw [opacity=0] (-1.3,-4.3)--(1.7,-1.3) node [opacity=0.5,sloped,midway] {$R_2+2R_3\geq \frac{1}{2}H_2+H_3$};
		
		\node [opacity=.5, right](R13) at (1.5,0.2) {$R_1+R_3\geq\frac{1}{2} H_2+H_3$};
		\draw [opacity=0] (0.5,0.5)--(0.5,4.5) node [opacity=0.5,sloped,midway] {$R_1+R_2\geq \frac{1}{2}H_2+H_3$};
		\draw [opacity=0] (-3.5,-4.0)--(0.5,0) node [opacity=0.5,sloped,midway] {$R_2+R_3\geq \frac{1}{2}H_2+H_3$};
		\end{tikzpicture}
		\caption{rate region $\cR_1^*$: case i ({\scriptsize $H_2<\frac{2}{3}H_3$})}
		\label{fig-region-case1}
	\end{figure}
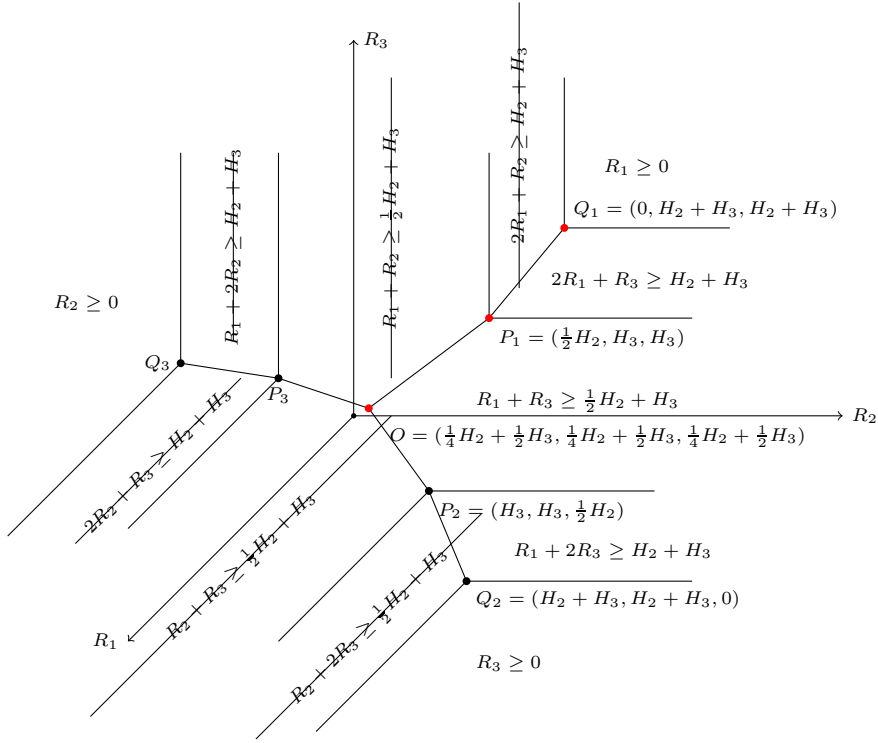

	The set of conditions $m=\frac{1}{2}H(X_2)+H(X_3)$, $\frac{3}{2}m\geq\frac{3}{2}H(X_2)+H(X_3)$, and $2m\geq 2H(X_2)+H(X_3)$ are equivalent to the constraint 
	\begin{equation}
	H(X_2)<\frac{2}{3}H(X_3). \label{constaint-case1}
	\end{equation}
	We can check that, under the condition in \eqref{constaint-case1}, the second constraint \eqref{rate-mss32-2-1} of $\cR_1^*$ implies the last two \eqref{rate-mss32-3}\eqref{rate-mss32-4}, thus the last two constraints are inactive.
	The rate region $\cR_1^*$ becomes the set of nonnegative rate tuples $(R_1,R_2,R_3)$ such that
	\begin{align}
	2R_i+R_j&\geq H(X_2)+H(X_3), \text{ for }1\leq i,j\leq 3 \text{ and }i\neq j \label{case1-rate-mss32-1} \\
	R_i+R_j &\geq \frac{1}{2}H(X_2)+H(X_3), \text{ for }1\leq i<j\leq 3.  \label{case1-rate-mss32-2}
	\end{align}
	The rate region is drawn in \reffig{fig-region-case1}.
	For abbreviation, we denote $H(X_2)$ and $H(X_3)$ by $H_2$ and $H_3$ for labeling in the figures. For sufficiently large $n$, the lengths $l_2$ and $l_3$ are large enough so that we can alway asymptotically partition the new source sequences $\bm{X}_2$ and $\bm{X}_3$ properly.
	
	Due to time sharing arguments and symmetry, we only need to show the achievability for the corner points $Q_1=(0,H_2+H_3,H_2+H_3)$, $P_1=(\frac{1}{2}H_2,H_3,H_3)$ and $O=(\frac{1}{4}H_2+\frac{1}{2}H_3, \frac{1}{4}H_2+\frac{1}{2}H_3, \frac{1}{4}H_2+\frac{1}{2}H_3)$. 
	\begin{enumerate}
		\item corner point $Q_1$: Let the secret key $\bm{Z}_1$ be a random binary sequence with the same length as $\bm{X}_3$.we use the following scheme which works for all $H_2$ and $H_3$ (in particular for $H_2<\frac{2}{3}H_3$)
		\begin{equation}
		\begin{cases}
		W_1=\emptyset \\
		W_2=(\bm{X}_2,\bm{Z}_1) \\
		W_3=(\bm{X}_2,\bm{X}_3+\bm{Z}_1).
		\end{cases} \label{scheme1-Q1}
		\end{equation}
		It is easy to check that
		\begin{equation}
		(R_1,R_2,R_3)=(0,H_2+H_3,H_2+H_3)
		\end{equation}
		and any two channels can losslessly reconstruct $\bm{X}_2$, all the three channels can reconstruct $\bm{X}_3$, and any one channel know nothing (information theoretic) about $\bm{X}_3$. We will only describe the coding schemes and omit writing out such checks in the sequel.
		
		\item corner point $P_1$: equipartition $\bm{X}_2$ into $\bm{A}_2$ and $\bm{B}_2$ with the same length $\frac{1}{2}l_2$. Partition $\bm{X}_3$ into $\bm{A}_3$, $\bm{B}_3$, and $\bm{C}_3$ with lengths $\frac{1}{2}l_2$, $\frac{1}{2}l_2$, and $l_3-l_2$. The length $l_3-l_2$ is nonnegative for all $H_2$ and $H_3$ such that $H_2\leq H_3$, in particular for $H_2<\frac{2}{3}H_3$. Let $\bm{Z}_2$ be a random binary sequence with the same length as $\bm{C}_3$. Then we use the following scheme which works for all $H_2$ and $H_3$ such that $H_2\leq H_3$ (in particular for $H_2<\frac{2}{3}H_3$):
		\begin{equation}
		\begin{cases}
		W_1=(\bm{A}_2+\bm{B}_2) \\
		W_2=(\quad~\bm{A}_2,\quad~\bm{A}_3+\bm{B}_2,\qquad~\bm{Z}_2) \\
		W_3=(\quad~\bm{B}_2,\quad~\bm{B}_3+\bm{A}_2,\quad~\bm{C}_3+\bm{Z}_2).
		\end{cases} \label{scheme1-P1}
		\end{equation}
		We can check that
		\begin{equation}
		(R_1,R_2,R_3)=(\frac{1}{2}H_2,H_3,H_3).
		\end{equation}
		
		\item corner point $O$: equipartition $\bm{X}_2$ into $\bm{A}_2$ and $\bm{B}_2$ with the same length $\frac{1}{2}l_2$. Partition $\bm{X}_3$ into $\bm{A}_3$, $\bm{B}_3$, $\bm{C}_3$, $\bm{D}_3$, and $\bm{E}_3$ with lengths $\frac{1}{2}l_2$, $\frac{1}{2}l_2$, $\frac{1}{2}l_2$, $\frac{1}{2}l_3-\frac{3}{4}l_2$, and $\frac{1}{2}l_3-\frac{3}{4}l_2$. The length $\frac{1}{2}l_3-\frac{3}{4}l_2$ is positive since $H_2<\frac{2}{3}H_3$. Let $\bm{Z}_3$ be a random binary sequence with the same length as $\bm{D}_3$ and $\bm{E}_3$. Then we use the following scheme:
		\begin{equation}
		\begin{cases}
		W_1=(\bm{A}_2+\bm{B}_2,\quad \bm{A}_3+\bm{A}_2,\qquad~\bm{Z}_3) \\
		W_2=(\quad~\bm{A}_2,\qquad~\bm{B}_3+\bm{B}_2,\quad~\bm{D}_3+\bm{Z}_3) \\
		W_3=(\quad~\bm{B}_2,\qquad~\bm{C}_3+\bm{A}_2,\quad~\bm{E}_3+\bm{Z}_3).
		\end{cases} \label{scheme1-O}
		\end{equation}
		We can check that
		\begin{align}
		R_1=R_2=R_3&=\frac{1}{2}H_2+\frac{1}{2}H_2+(\frac{1}{2}H_3-\frac{3}{4}H_2)  \nonumber \\
		&=\frac{1}{4}H_2+\frac{1}{2}H_3.
		\end{align}
	\end{enumerate}
	Therefore, we finish the achievability proof for the case that $H_2<\frac{2}{3}H_3$.
	
	\subsection{Case ii}  \label{proof-achievability-thm32mss-case2}
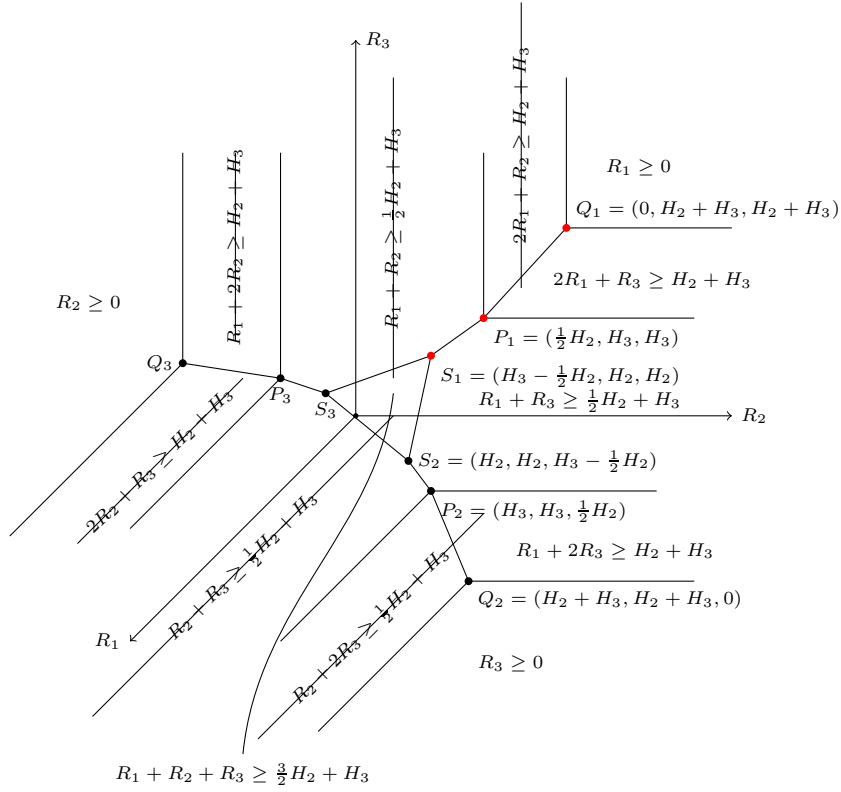
\begin{figure}[!h]
		\centering
		\begin{tikzpicture}[font=\scriptsize, scale=1.0]
		\draw [->,opacity=0.2] (0,0)--(-3.0,-3.0) node [left,black,opacity=.5](R1) {$R_1$};
		\draw [->,opacity=0.2] (0,0)--(5.0,0) node [right,black,opacity=.5](R2) {$R_2$};
		\draw [->,opacity=0.2] (0,0)--(0,5.0) node [right,black,opacity=.5](R3) {$R_3$};
		\fill [black,opacity=.5] (0,0) circle (1.0pt);
		
		\coordinate (Q1) at (2.8,2.5);
		\coordinate (Q2) at (1.5,-2.2);
		\coordinate (Q3) at (-2.3,0.7);
		\coordinate (P1) at (1.7,1.3);
		\coordinate (P2) at (1.0,-1.0);
		\coordinate (P3) at (-1.0,0.5);
		\coordinate (S1) at (1.0,0.8);
		\coordinate (S2) at (0.7,-0.6);
		\coordinate (S3) at (-0.4,0.3);
		
		\coordinate (Q1') at (2.8,4.5);
		\coordinate (Q1'') at (5.0,2.5);
		\coordinate (Q2') at (4.5,-2.2);
		\coordinate (Q2'') at (-0.5,-4.2);
		\coordinate (Q3') at (-4.6,-1.6);
		\coordinate (Q3'') at (-2.3,3.5);
		
		\coordinate (P1') at (4.5,1.3);
		\coordinate (P1'') at (1.7,3.5);
		\coordinate (P2') at (4.0,-1.0);
		\coordinate (P2'') at (-1.0,-3.0);
		\coordinate (P3') at (-1.0,3.5);
		\coordinate (P3'') at (-3.0,-1.5);
		
		\draw (S1)--(S2)--(S3)--cycle;
		\foreach \j in {1,2,3}{\draw (Q\j)--(Q\j') (Q\j)--(Q\j'') (P\j)--(P\j') (P\j)--(P\j'') (Q\j)--(P\j)--(S\j);}
		
		\foreach \point in {Q2,Q3,S2,S3,P2,P3}{\fill [black,opacity=.5] (\point) circle (1.5pt);}
		\foreach \point in {Q1,P1,S1}{\fill [red,opacity=.5] (\point) circle (1.5pt);}
		
		\node [above right](Q1) at (Q1) {$Q_1=(0,H_2+H_3,H_2+H_3)$};
		\node [below right](Q2) at (Q2) {$Q_2=(H_2+H_3,H_2+H_3,0)$};
		\node [left](Q3) at (Q3) {$Q_3$};
		\node [below right](S1) at (S1) {$S_1=(H_3-\frac{1}{2}H_2,H_2,H_2)$};
		\node [ right](S2) at (S2) {$S_2=(H_2,H_2,H_3-\frac{1}{2}H_2)$};
		\node [below](S3) at (S3)  {$S_3$};
		\node [below right](P1) at (P1) {$P_1=(\frac{1}{2}H_2,H_3,H_3)$};
		\node [below right](P2) at (P2) {$P_2=(H_3,H_3,\frac{1}{2}H_2)$};
		\node [below](P3) at (P3)  {$P_3$};
		
		\node [opacity=.5, right](R1) at (3.2,3.3) {$R_1\geq 0$};
		\node [opacity=.5, right](R3) at (1.5,-3.3) {$R_3\geq 0$};
		\node [opacity=.5, left](R2) at (-3.0,1.5) {$R_2\geq 0$};
		
		\node [opacity=.5, right](2R1R3) at (2.5,1.8) {$2R_1+R_3\geq H_2+H_3$};
		\node [opacity=.5, right](2R3R1) at (2.0,-1.8) {$R_1+2R_3\geq H_2+H_3$};
		\draw [opacity=0] (2.2,1.7)--(2.2,5.5) node [opacity=0.5,sloped,midway] {$2R_1+R_2\geq H_2+H_3$};
		\draw [opacity=0] (-1.6,1.0)--(-1.6,3.5) node [opacity=0.5,sloped,midway] {$R_1+2R_2\geq H_2+H_3$};
		\draw [opacity=0] (-3.7,-1.7)--(-1.5,0.5) node [opacity=0.5,sloped,midway] {$2R_2+R_3\geq H_2+H_3$};
		\draw [opacity=0] (-1.3,-4.3)--(1.7,-1.3) node [opacity=0.5,sloped,midway] {$R_2+2R_3\geq \frac{1}{2}H_2+H_3$};
		
		\node [opacity=.5, right](R13) at (1.5,0.2) {$R_1+R_3\geq\frac{1}{2} H_2+H_3$};
		\draw [opacity=0] (0.5,0.5)--(0.5,4.5) node [opacity=0.5,sloped,midway] {$R_1+R_2\geq \frac{1}{2}H_2+H_3$};
		\draw [opacity=0] (-3.5,-4.0)--(0.5,0) node [opacity=0.5,sloped,midway] {$R_2+R_3\geq \frac{1}{2}H_2+H_3$};
		
		\draw [opacity=0.2] (0.5,0.3) .. controls (0.3,-1.5) and (-1.3,-2.5) .. (-1.5,-4.5) node[opacity=.5, below]{$R_1+R_2+R_3\geq \frac{3}{2}H_2+H_3$};
		\end{tikzpicture}
		\caption{rate region $\cR_1^*$: case ii ({\scriptsize $\frac{2}{3}H_3\leq H_2<H_3$})}
		\label{fig-region-case2}
	\end{figure}

	The set of conditions $m=\frac{1}{2}H(X_2)+H(X_3)$, $\frac{3}{2}m\leq\frac{3}{2}H(X_2)+H(X_3)$, and $2m> 2H(X_2)+H(X_3)$ are equivalent to the constraint 
	\begin{equation}
	\frac{2}{3}H(X_3)\leq H(X_2)<H(X_3). \label{constaint-case2}
	\end{equation}
	We can check that, under the condition in \eqref{constaint-case2}, the second constraint \eqref{rate-mss32-2-1} of $\cR_1^*$ implies the last one \eqref{rate-mss32-4}, thus the last constraint is inactive.
	The rate region $\cR_1^*$ becomes the set of nonnegative rate tuples $(R_1,R_2,R_3)$ such that
	\begin{align}
	2R_i+R_j&\geq H(X_2)+H(X_3), \text{ for }1\leq i,j\leq 3 \text{ and }i\neq j \label{case2-rate-mss32-1} \\
	R_i+R_j &\geq \frac{1}{2}H(X_2)+H(X_3), \text{ for }1\leq i<j\leq 3  \label{case2-rate-mss32-2}  \\
	R_1+R_2+R_3&\geq \frac{3}{2}H(X_2)+H(X_3). \label{case2-rate-mss32-3}
	\end{align}
	The rate region is depicted in \reffig{fig-region-case2}.
	Due to time-sharing arguments and symmetry, we only need to show the achievability for the corner points $Q_1=(0,H_2+H_3,H_2+H_3)$, $P_1=(\frac{1}{2}H_2,H_3,H_3)$ and $S_1=(H_3-\frac{1}{2}H_2,H_2,H_2)$. 
	\begin{enumerate}
		\item corner point $Q_1$: the same as $Q_1$ in \eqref{scheme1-Q1}
		
		\item corner point $P_1$: the same as $P_1$ in \eqref{scheme1-P1}
		
		\item corner point $S_1$: equipartition $\bm{X}_2$ into $\bm{A}_2$ and $\bm{B}_2$ with the same length $\frac{1}{2}l_2$. Partition $\bm{X}_3$ into $\bm{A}_3$, $\bm{B}_3$, and $\bm{C}_3$ with lengths $\frac{1}{2}l_2$, $\frac{1}{2}l_2$, and $l_3-l_2$. The condition $H_2<H_3$ implies that the length $l_3-l_2$ is positive. Since $\frac{2}{3}H_3\leq H_2$ implies $l_3-l_2\leq \frac{1}{2}l_2$, we denote the first $l_3-l_2$ bits of $\bm{A}_2$ by $\bm{A}_2^1$. Then we use the following scheme:
		\begin{equation}
		\begin{cases}
		W_1=(\bm{A}_2+\bm{B}_2, \quad~\bm{C}_3+\bm{A}_2^1) \\
		W_2=(\quad~\bm{A}_2, \qquad~\bm{A}_3+\bm{B}_2) \\
		W_3=(\quad~\bm{B}_2, \qquad~\bm{B}_3+\bm{A}_2).
		\end{cases} \label{scheme2-S1}
		\end{equation}
		It is easy to check that
		\begin{align}
		R_1&=\frac{1}{2}H_2+(H_3-H_2)=H_3-\frac{1}{2}H_2 \\
		R_2=R_3&=\frac{1}{2}H_2+\frac{1}{2}H_2=H_2.
		\end{align}
		It may not be so obvious to see that $(\bm{A}_2+\bm{B}_2,~\bm{C}_3+\bm{A}_2^1)$ provide no information about $\bm{C}_3$. 
		So we state the result in the following lemma, which ensures that $W_1$ provide no information about $\bm{X}_3$. 
		The assumption that $\bm{X}_2$ and $\bm{X}_3$ are memoryless symmetric sources indicates all the bits of $\bm{X}_2$ and $\bm{X}_3$ are linearly independent. 
		Since the addition operations $\bm{A}_2+\bm{B}_2,~\bm{C}_3+\bm{A}_2^1$ are bitwise, we only state the claim for single bits $A$, $B$, and $C$ in the lemma.
		\begin{lemma}
			Let $A$, $B$, and $C$ be three independent random variables taking values in the same finite field $\mathbb{F}_2$. If $X=A+C$ and $Y=A+B$, then 
			\begin{equation}
			H(C|X,Y)=H(C).
			\end{equation}
		\end{lemma}
		\begin{proof}
			We try to show that $I(X,Y;C)=0$. Since mutual information is nonnegative, we only need to prove that $I(X,Y;C)\leq 0$. We have the following.
			\begin{align}
			-I(X,Y;C)&=H(C,X,Y)-H(X,Y)-H(C)  \nonumber \\
			&=\big[I(A;Y|C,X)+H(A,C,X,Y)-H(A,C,X)+H(C,X)\big] \nonumber \\
			&\quad +\big[I(X;Y)-H(X)-H(Y)\big]-H(C)   \nonumber \\
			&=I(A;Y|C,X)+H(A,C,Y)-H(A,C)+\big[H(C)+H(X)\big] \nonumber \\
			&\quad +I(X;Y)-H(X)-H(Y)-H(C)  \label{secret-1} \\
			&=I(A;Y|C,X)+I(X;Y)+H(A,C,Y)-H(A,C)-H(Y) \nonumber \\
			&=I(A;Y|C,X)+I(X;Y)-H(A,C)-H(Y)  \nonumber \\
			&\quad +\big[I(B;C|A,Y)+H(A,B,C,Y)-H(A,B,Y)+H(A,Y)\big]  \nonumber \\
			&=I(A;Y|C,X)+I(X;Y)-H(A,C)-H(Y)  \nonumber \\
			&\quad +I(B;C|A,Y)+H(A,B,C)-H(A,B)+\big[H(A)+H(Y)\big]  \label{secret-2}\\
			&=I(A;Y|C,X)+I(X;Y)+I(B;C|A,Y)  \nonumber \\
			&\quad +H(A,B,C)+H(A)-H(A,B)-H(A,C) \nonumber \\
			&=I(A;Y|C,X)+I(X;Y)+I(B;C|A,Y)  \label{secret-3}  \\
			&\geq 0,
			\end{align}
			where \eqref{secret-1} follows from 
			\begin{equation}
			H(X|A,C)=0
			\end{equation}
			and 
			\begin{equation}
			H(C,X)=H(C)+H(X),
			\end{equation}
			\eqref{secret-2} follows from 
			\begin{equation}
			H(Y|A,B)=0
			\end{equation}
			and
			\begin{equation}
			H(A,Y)=H(A)+H(Y),
			\end{equation}
			and \eqref{secret-3} follows from the mutual independence of $A$, $B$, and $C$. Then we conclude that $I(X,Y;C)=0$ and thus $H(C|X,Y)=H(C)$.
		\end{proof}
		
	\end{enumerate}
	Therefore, the achievability for the case that $\frac{2}{3}H_3\leq H_2<H_3$ is proved.
	
	\subsection{Case iii} \label{proof-achievability-thm32mss-case3}
\begin{figure}[!h]
		\centering
		\begin{tikzpicture}[font=\scriptsize, scale=1.0]
		\draw [->,opacity=0.2] (0.8,0)--(-2.2,-3.0) node [left,black,opacity=.5](R1) {$R_1$};
		\draw [->,opacity=0.2] (0.8,0)--(6.0,0) node [right,black,opacity=.5](R2) {$R_2$};
		\draw [->,opacity=0.2] (0.8,0)--(0.8,5.0) node [right,black,opacity=.5](R3) {$R_3$};
		\fill [black,opacity=.5] (0.8,0) circle (1.0pt);
		
		\coordinate (Q1) at (3.4,2.8);
		\coordinate (Q2) at (2.3,-3.2);
		\coordinate (Q3) at (-2.8,1.0);
		\coordinate (T1) at (2.8,2.2);
		\coordinate (T2) at (2.0,-2.3);
		\coordinate (T3) at (-1.9,0.8);
		\coordinate (S4) at (2.1,0.8);
		\coordinate (S5) at (1.8,-0.8);
		\coordinate (S6) at (1.0,-1.0);
		\coordinate (S7) at (-0.4,0.2);
		\coordinate (S8) at (-0.2,0.7);
		\coordinate (S9) at (1.5,1.25);
		
		\coordinate (Q1') at (3.4,4.5);
		\coordinate (Q1'') at (5.5,2.8);
		\coordinate (Q2') at (5.5,-3.2);
		\coordinate (Q2'') at (0.3,-4.9);
		\coordinate (Q3') at (-4.6,-0.8);
		\coordinate (Q3'') at (-2.8,4.0);
		\coordinate (S4') at (5.0,0.8);
		\coordinate (S5') at (5.0,-0.8);
		\coordinate (S6') at (-0.8,-2.8);
		\coordinate (S7') at (-2.3,-1.7);
		\coordinate (S8') at (-0.2,3.7);
		\coordinate (S9') at (1.5,4.0);
		
		\draw (Q1)--(T1) (Q2)--(T2) (Q3)--(T3);
		\draw (T1)--(S4)--(S5)--(T2)--(S6)--(S7)--(T3)--(S8)--(S9)--(T1)--cycle;
		\draw (S4')--(S4)--(S9)--(S9') (S5')--(S5)--(S6)--(S6') (S7')--(S7)--(S8)--(S8');
		
		\foreach \j in {1,2,3}{
			\draw (Q\j)--(Q\j') (Q\j)--(Q\j'');}
		
		\foreach \point in {Q2,Q3,T2,T3,S5,S6,S7,S8,S9}
		{\fill [black,opacity=.4] (\point) circle (1.5pt);}
		\foreach \point in {Q1,T1,S4}
		{\fill [red,opacity=.5] (\point) circle (1.5pt);}
		
		\node [above right](Q1) at (Q1) {$Q_1=(0,H_2+H_3,H_2+H_3)$};
		\node [below right](Q2) at (Q2) {$Q_2=(H_2+H_3,H_2+H_3,0)$};
		\node [left](Q3) at (Q3) {$Q_3$};
		\node [right](T1) at (T1) {$T_1=(\frac{1}{2}H_3,H_2,H_2)$};
		\node [below right](T2) at (T2) {$T_2=(H_2,H_2,\frac{1}{2}H_3)$};
		\node [above](T3) at (T3) {$T_3$};
		\node [below right](S4) at (S4) {$S_4=(\frac{1}{2}H_2,H_2,H_3)$};
		\node [above right](S5) at (S5) {$S_5=(H_3,H_2,\frac{1}{2}H_2)$};
		\node [above](S6) at (S6) {$S_6$};
		\node [right](S7) at (S7) {$S_7$};
		\node [above right](S8) at (S8) {$S_8$};
		\node [below](S9) at (S9) {$S_9$};
		
		\node [opacity=.5, right](R1) at (3.8,3.8) {$R_1\geq 0$};
		\node [opacity=.5, left](R2) at (-3.0,1.5) {$R_2\geq 0$};
		\node [opacity=.5, right](R3) at (2.5,-4.3) {$R_3\geq 0$};
		
		\node [opacity=.5, right](2R1R3) at (3.0,1.8) {$2R_1+R_3\geq H_2+H_3$};
		\node [opacity=.5, right](2R3R1) at (2.5,-2.0) {$R_1+2R_3\geq H_2+H_3$};
		\draw [opacity=0] (2.2,1.7)--(2.2,5.5) node [opacity=0.5,sloped,midway] {$2R_1+R_2\geq H_2+H_3$};
		\draw [opacity=0] (-1.7,1.0)--(-1.7,4.0) node [opacity=0.5,sloped,midway] {$R_1+2R_2\geq H_2+H_3$};
		\draw [opacity=0] (-3.7,-1.7)--(-1.5,0.5) node [opacity=0.5,sloped,midway] {$2R_2+R_3\geq H_2+H_3$};
		\draw [opacity=0] (-1.3,-4.3)--(1.7,-1.3) node [opacity=0.5,sloped,midway] {$R_2+2R_3\geq \frac{1}{2}H_2+H_3$};
		
		\draw [opacity=0] (0.5,0.5)--(0.5,4.5) node [opacity=0.5,sloped,midway] {$R_1+R_2\geq \frac{1}{2}H_2+H_3$};
		\node [opacity=.5, right](R13) at (2.5,0.2) {$R_1+R_3\geq\frac{1}{2} H_2+H_3$};
		\draw [opacity=0] (-2.5,-3.0)--(0.5,0) node [opacity=0.5,sloped,midway] {$R_2+R_3\geq \frac{1}{2}H_2+H_3$};
		
		\draw [opacity=0.2] (1.2,0.4) .. controls (0.4,-1.5) and (-1.5,-2.5) .. (-1.5,-4.5) node[opacity=.5, below]{$R_1+R_2+R_3\geq \frac{3}{2}H_2+H_3$};
		
		\draw [opacity=0.2] (2.0,1.3)--(3.3,1.3) node[opacity=.5, right]{$2R_1+R_2+R_3\geq 2H_2+H_3$};
		\draw [opacity=0.2] (1.5,-1.3)--(2.8,-1.3) node[opacity=.5, right]{$R_1+R_2+2R_3\geq 2H_2+H_3$};
		\draw [opacity=0.2] (-0.8,0.5)--(-0.8,1.3); \draw [opacity=0] (-0.8,0.5)--(-0.8,3.5) node[opacity=.5, sloped, very near end]{$R_1+2R_2+R_3\geq 2H_2+H_3$};
		\end{tikzpicture}
		\caption{rate region $\cR_1^*$: case iii ({\scriptsize $H_3\leq H_2<2H_3$})}
		\label{fig-region-case3}
	\end{figure}
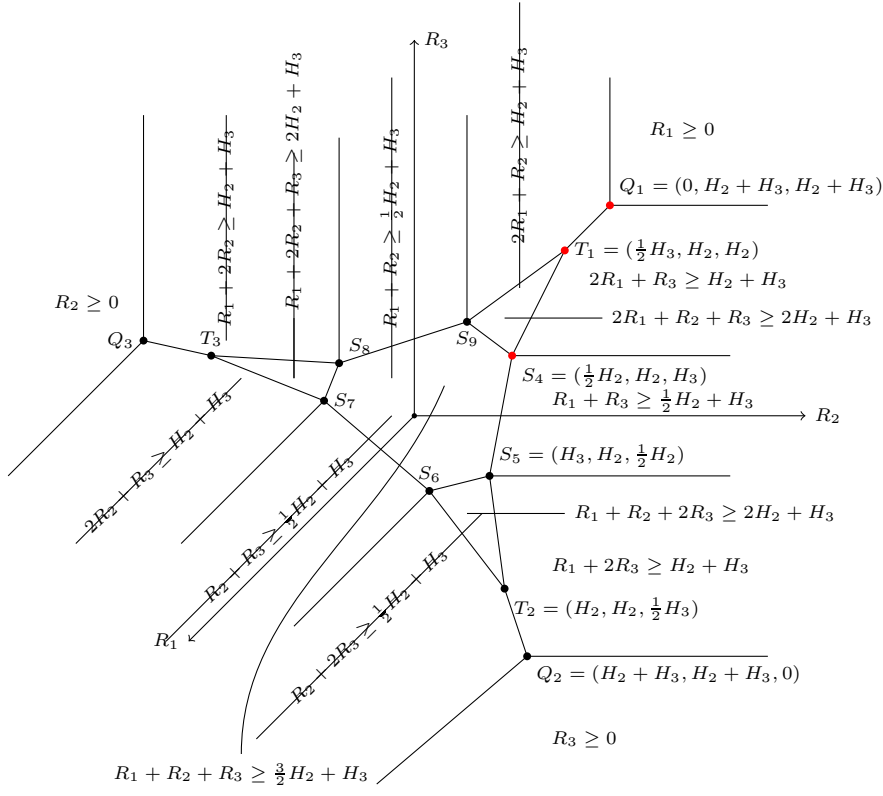

	The set of conditions $m=\frac{1}{2}H(X_2)+H(X_3)$, $\frac{3}{2}m\leq\frac{3}{2}H(X_2)+H(X_3)$, and $2m\leq 2H(X_2)+H(X_3)$ are equivalent to the constraint 
	\begin{equation}
	H(X_3)\leq H(X_2)<2H(X_3). \label{constaint-case3}
	\end{equation}
	We can see that, under the condition in \eqref{constaint-case3}, all the constraints in \eqref{rate-mss32-1}-\eqref{rate-mss32-4} are active. The rate region $\cR_1^*$ becomes the set of nonnegative rate tuples $(R_1,R_2,R_3)$ such that
	\begin{align}
	2R_i+R_j&\geq H(X_2)+H(X_3), \text{ for }1\leq i,j\leq 3 \text{ and }i\neq j \label{case3-rate-mss32-1} \\
	R_i+R_j &\geq \frac{1}{2}H(X_2)+H(X_3), \text{ for }1\leq i<j\leq 3  \label{case3-rate-mss32-2}  \\
	R_1+R_2+R_3&\geq \frac{3}{2}H(X_2)+H(X_3) \label{case3-rate-mss32-3} \\
	2R_i+R_{i\odot 1}+R_{i\odot 2}&\geq 2H(X_2)+H(X_3), \text{ for }1\leq i\leq 3. \label{case3-rate-mss32-4}
	\end{align}
	The rate region is depicted in \reffig{fig-region-case3}.
	Due to time-sharing arguments and symmetry, we only need to show the achievability for the corner points $Q_1=(0,H_2+H_3,H_2+H_3)$, $T_1=(\frac{1}{2}H_3,H_2,H_2)$ and $S_4=(\frac{1}{2}H_2,H_2,H_3)$. 
	\begin{enumerate}
		\item corner point $Q_1$: the same as $Q_1$ in \eqref{scheme1-Q1}
		
		\item corner point $T_1$: partition $\bm{X}_2$ into $\bm{A}_2$, $\bm{B}_2$, and $\bm{C}_2$ with lengths $\frac{1}{2}l_3$, $\frac{1}{2}l_3$, and $l_2-l_3$. The length $l_2-l_3$ is nonnegative for all pairs $(H_2,H_3)$ such that $H_3\leq H_2$, in particular for $H_3\leq H_2<2H_3$. Equipartition $\bm{X}_3$ into $\bm{A}_3$ and $\bm{B}_3$ with the same length $\frac{1}{2}l_3$. Then we use the following scheme which works for all $H_2$ and $H_3$ such that $H_3\leq H_2$ (in particular for $H_3\leq H_2<2H_3$):
		\begin{equation}
		\begin{cases}
		W_1=(\bm{A}_2+\bm{B}_2) \\
		W_2=(\quad~ \bm{A}_2,\quad~\bm{C}_2,\quad~\bm{A}_3+\bm{B}_2) \\
		W_3=(\quad~\bm{B}_2,\quad~\bm{C}_2,\quad~\bm{B}_3+\bm{A}_2).
		\end{cases} \label{scheme3-T1}
		\end{equation}
		We can check that 
		\begin{equation}
		(R_1,R_2,R_3)=(\frac{1}{2}H_3,H_2,H_2).
		\end{equation}
		
		\item corner point $S_4$: equipartition $\bm{X}_2$ into $\bm{A}_2$ and $\bm{B}_2$ with the same length $\frac{1}{2}l_2$. Partition $\bm{X}_3$ into $\bm{A}_3$ and $\bm{B}_3$ with lengths $\frac{1}{2}l_2$ and $l_3-\frac{1}{2}l_2$. Since $H_3\leq H_2$ implies $l_3-\frac{1}{2}l_2\leq \frac{1}{2}l_2$, we denote the first $l_3-\frac{1}{2}l_2$ bits of $\bm{A}_2$ by $\bm{A}_2^1$. Then we use the following scheme:
		\begin{equation}
		\begin{cases}
		W_1=(\bm{A}_2+\bm{B}_2) \\
		W_2=(\quad~ \bm{A}_2,\quad~ \bm{A}_3+\bm{B}_2) \\
		W_3=(\quad~ \bm{B}_2,\quad~ \bm{B}_3+\bm{A}_2^1).
		\end{cases} \label{scheme3-S4}
		\end{equation}
		It's easy to check that 
		\begin{equation}
		(R_1,R_2,R_3)=(\frac{1}{2}H_2,H_2,H_3).
		\end{equation}
	\end{enumerate}
	Therefore, we prove the achievability for the case that $H(X_3)\leq H(X_2)<2H(X_3)$.
	
	\subsection{Case iv}  \label{proof-achievability-thm32mss-case4}
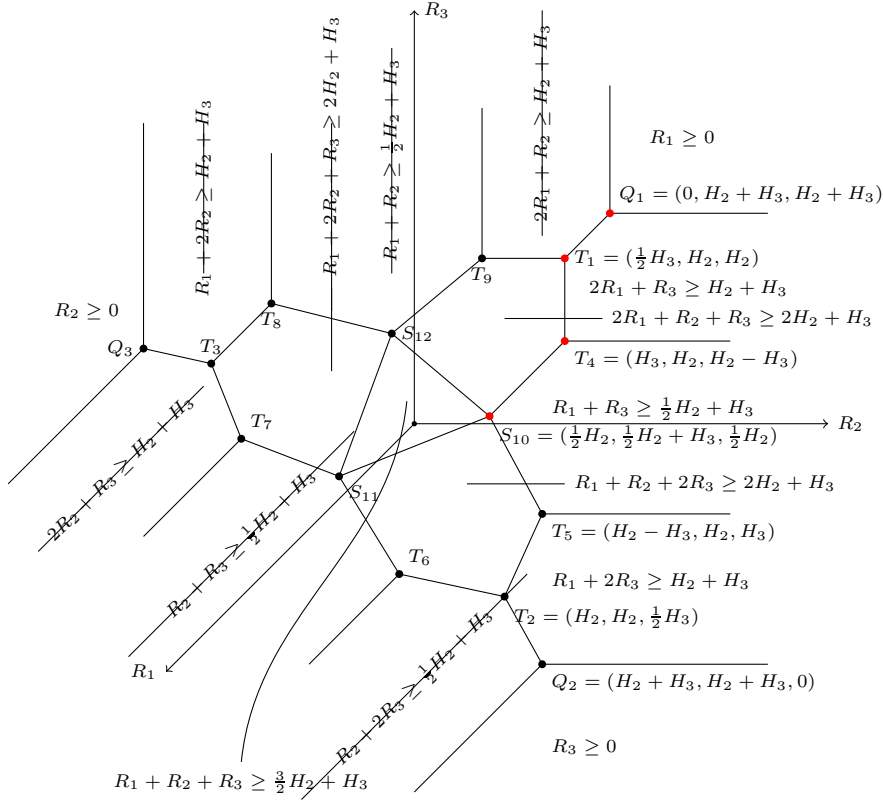
\begin{figure}[!h]
		\centering
		\begin{tikzpicture}[font=\scriptsize, scale=1.0]
		\draw [->,opacity=0.2] (0.8,0)--(-2.5,-3.3) node [left,black,opacity=.5](R1) {$R_1$};
		\draw [->,opacity=0.2] (0.8,0)--(6.3,0) node [right,black,opacity=.5](R2) {$R_2$};
		\draw [->,opacity=0.2] (0.8,0)--(0.8,5.5) node [right,black,opacity=.5](R3) {$R_3$};
		\fill [black,opacity=.5] (0.8,0) circle (1.0pt);
		
		\coordinate (Q1) at (3.4,2.8);
		\coordinate (Q2) at (2.5,-3.2);
		\coordinate (Q3) at (-2.8,1.0);
		\coordinate (T1) at (2.8,2.2);
		\coordinate (T2) at (2.0,-2.3);
		\coordinate (T3) at (-1.9,0.8);
		\coordinate (T4) at (2.8,1.1);
		\coordinate (T5) at (2.5,-1.2);
		\coordinate (T6) at (0.6,-2.0);
		\coordinate (T7) at (-1.5,-0.2);
		\coordinate (T8) at (-1.1,1.6);
		\coordinate (T9) at (1.7,2.2);
		
		\coordinate (S10) at (1.8,0.1);
		\coordinate (S11) at (-0.2,-0.7);
		\coordinate (S12) at (0.5,1.2);
		
		\coordinate (Q1') at (3.4,4.5);
		\coordinate (Q1'') at (5.5,2.8);
		\coordinate (Q2') at (5.5,-3.2);
		\coordinate (Q2'') at (0.8,-4.9);
		\coordinate (Q3') at (-4.6,-0.8);
		\coordinate (Q3'') at (-2.8,4.0);
		\coordinate (T4') at (5.0,1.1);
		\coordinate (T5') at (5.0,-1.2);
		\coordinate (T6') at (-0.6,-3.2);
		\coordinate (T7') at (-2.8,-1.5);
		\coordinate (T8') at (-1.1,3.6);
		\coordinate (T9') at (1.7,4.2);
		
		\draw (S10)--(S11)--(S12)--(S10);
		\draw (Q1)--(T1) (Q2)--(T2) (Q3)--(T3);
		\draw (T1)--(T4)--(S10)--(T5)--(T2)--(T6)--(S11)--(T7)--(T3)--(T8)--(S12)--(T9)--(T1)--cycle;
		
		\foreach \j in {4,5,6,7,8,9}{\draw (T\j)--(T\j');}
		\foreach \j in {1,2,3}{\draw (Q\j)--(Q\j') (Q\j)--(Q\j'');}
		
		\foreach \point in {Q2,Q3,T2,T3,T5,T6,T7,T8,T9,S11,S12}
		{\fill [black,opacity=.4] (\point) circle (1.5pt);}
		\foreach \point in {Q1,T1,T4,S10}
		{\fill [red,opacity=.5] (\point) circle (1.5pt);}
		
		\node [above right](Q1) at (Q1) {$Q_1=(0,H_2+H_3,H_2+H_3)$};
		\node [below right](Q2) at (Q2) {$Q_2=(H_2+H_3,H_2+H_3,0)$};
		\node [left](Q3) at (Q3) {$Q_3$};
		\node [right](T1) at (T1) {$T_1=(\frac{1}{2}H_3,H_2,H_2)$};
		\node [below right](T2) at (T2) {$T_2=(H_2,H_2,\frac{1}{2}H_3)$};
		\node [above](T3) at (T3) {$T_3$};
		\node [below right](T4) at (T4) {$T_4=(H_3,H_2,H_2-H_3)$};
		\node [below right](T5) at (T5) {$T_5=(H_2-H_3,H_2,H_3)$};
		\node [above right](T6) at (T6) {$T_6$};
		\node [above right](T7) at (T7) {$T_7$};
		\node [below](T8) at (T8) {$T_8$};
		\node [below](T9) at (T9) {$T_9$};
		\node [below right](S10) at (S10) {$S_{10}=(\frac{1}{2}H_2,\frac{1}{2}H_2+H_3,\frac{1}{2}H_2)$};
		\node [below right](S11) at (S11) {$S_{11}$};
		\node [right](S12) at (S12) {$S_{12}$};
		
		\node [opacity=.5, right](R1) at (3.8,3.8) {$R_1\geq 0$};
		\node [opacity=.5, left](R2) at (-3.0,1.5) {$R_2\geq 0$};
		\node [opacity=.5, right](R3) at (2.5,-4.3) {$R_3\geq 0$};
		
		\node [opacity=.5, right](2R1R3) at (3.0,1.8) {$2R_1+R_3\geq H_2+H_3$};
		\node [opacity=.5, right](2R3R1) at (2.5,-2.1) {$R_1+2R_3\geq H_2+H_3$};
		\draw [opacity=0] (2.5,2.5)--(2.5,5.5) node [opacity=0.5,sloped,midway] {$2R_1+R_2\geq H_2+H_3$};
		\draw [opacity=0] (-2.0,2.0)--(-2.0,4.0) node [opacity=0.5,sloped,midway] {$R_1+2R_2\geq H_2+H_3$};
		\draw [opacity=0] (-4.2,-1.7)--(-2.0,0.5) node [opacity=0.5,sloped,midway] {$2R_2+R_3\geq H_2+H_3$};
		\draw [opacity=0] (-0.7,-5.0)--(2.3,-2.0) node [opacity=0.5,sloped,midway] {$R_2+2R_3\geq \frac{1}{2}H_2+H_3$};
		
		\draw [opacity=0] (0.5,2)--(0.5,5) node [opacity=0.5,sloped,midway] {$R_1+R_2\geq \frac{1}{2}H_2+H_3$};
		\node [opacity=.5, right](R13) at (2.5,0.2) {$R_1+R_3\geq\frac{1}{2} H_2+H_3$};
		\draw [opacity=0] (-3.0,-3.1)--(0,-0.1) node [opacity=0.5,sloped,midway] {$R_2+R_3\geq \frac{1}{2}H_2+H_3$};
		
		\draw [opacity=0.2] (0.7,0.3) .. controls (0.5,-1.7) and (-1.3,-2.5) .. (-1.5,-4.5) node[opacity=.5, below]{$R_1+R_2+R_3\geq \frac{3}{2}H_2+H_3$};
		
		\draw [opacity=0.2] (2.0,1.4)--(3.3,1.4) node[opacity=.5, right]{$2R_1+R_2+R_3\geq 2H_2+H_3$};
		\draw [opacity=0.2] (1.5,-0.8)--(2.8,-0.8) node[opacity=.5, right]{$R_1+R_2+2R_3\geq 2H_2+H_3$};
		\draw [opacity=0.2] (-0.3,0.7)--(-0.3,1.8); \draw [opacity=0] (-0.3,1.5)--(-0.3,4.0) node[opacity=.5, sloped, very near end]{$R_1+2R_2+R_3\geq 2H_2+H_3$};
		
		\end{tikzpicture}
		\caption{rate region $\cR_1^*$: case iv ({\scriptsize $2H_3\leq H_2$})}
		\label{fig-region-case4}
	\end{figure}

	The condition $m=H(X_2)$ is equivalent to the constraint 
	\begin{equation}
	2H(X_3)\leq H(X_2). \label{constaint-case4}
	\end{equation}
	We can see that, under the condition in \eqref{constaint-case4}, all the constraints in \eqref{rate-mss32-1}-\eqref{rate-mss32-4} are active. The rate region $\cR_1^*$ becomes the set of nonnegative rate tuples $(R_1,R_2,R_3)$ such that
	\begin{align}
	2R_i+R_j&\geq H(X_2)+H(X_3), \text{ for }1\leq i,j\leq 3 \text{ and }i\neq j \label{case4-rate-mss32-1} \\
	R_i+R_j &\geq H(X_2), \text{ for }1\leq i<j\leq 3  \label{case4-rate-mss32-2}  \\
	R_1+R_2+R_3&\geq \frac{3}{2}H(X_2)+H(X_3) \label{case4-rate-mss32-3} \\
	2R_i+R_{i\odot 1}+R_{i\odot 2}&\geq 2H(X_2)+H(X_3), \text{ for }1\leq i\leq 3. \label{case4-rate-mss32-4}
	\end{align}
	The rate region is depicted in \reffig{fig-region-case4}.
	Due to time-sharing arguments and symmetry, we only need to show the achievability for the corner points $Q_1=(0,H_2+H_3,H_2+H_3)$, $T_1=(\frac{1}{2}H_3,H_2,H_2)$, $T_4=(H_3,H_2,H_2-H_3)$ and $S_{10}=(\frac{1}{2}H_2,\frac{1}{2}H_2+H_3,\frac{1}{2}H_2)$. 
	\begin{enumerate}
		\item corner point $Q_1$: the same as $Q_1$ in \eqref{scheme1-Q1}
		\item corner point $T_1$: the same as $T_1$ in \eqref{scheme3-T1}
		
		\item corner point $T_4$: partition $\bm{X}_2$ into $\bm{A}_2$, $\bm{B}_2$, and $\bm{C}_2$ with lengths $l_3$, $l_3$, and $l_2-2l_3$. The length $l_2-2l_3$ is nonnegative since $2H_3\leq H_2$. Then we use the following scheme:
		\begin{equation}
		\begin{cases}
		W_1=(\bm{A}_2+\bm{B}_2) \\
		W_2=(\quad~ \bm{A}_2,\quad~\bm{C}_2,\quad~\bm{X}_3+\bm{B}_2) \\
		W_3=(\quad~\bm{B}_2,\quad~\bm{C}_2).
		\end{cases} \label{scheme4-T4}
		\end{equation}
		It's easy to check that
		\begin{equation}
		(R_1,R_2,R_3)=(H_3,H_2,H_2-H_3).
		\end{equation}
		
		\item corner point $S_{10}$: equipartition $\bm{X}_2$ into $\bm{A}_2$ and $\bm{B}_2$ with the same length $\frac{1}{2}l_2$. Since $2H_3\leq H_2$ implies $l_3\leq \frac{1}{2}l_2$, we denote the first $l_3$ bits of $\bm{B}_2$ by $\bm{B}_2^1$. Then we use the following scheme:
		\begin{equation}
		\begin{cases}
		W_1=(\bm{A}_2+\bm{B}_2) \\
		W_2=(\quad~ \bm{A}_2,\quad~ \bm{X}_3+\bm{B}_2^1) \\
		W_3=(\quad~ \bm{B}_2\quad~).
		\end{cases} \label{scheme4-S4}
		\end{equation}
		We can check that
		\begin{equation}
		(R_1,R_2,R_3)=(\frac{1}{2}H_2,\frac{1}{2}H_2+H_3,\frac{1}{2}H_2).
		\end{equation}
	\end{enumerate}
	Therefore, the achievability for the case that $2H(X_3)\leq H(X_2)$ is proved. 
	
	To sum up the four cases, we finish the achievability proof of \Cref{thm-32mss} for all $H(X_2)$ and $H(X_3)$.
	
	\section{Converse Proof of \Cref{thm-32mss}}  \label{proof-converse-thm32mss}
	From conditions in \eqref{recover constraint-mss32-2}-\eqref{security constraint-mss32}, we can write the entropies of $\bm{X}_2$ and $\bm{X}_3$ as follows.
	\begin{align}
	H(\bm{X}_2)&=I(\bm{X}_2;W_1W_2)+H(\bm{X}_2|W_1W_2) \nonumber \\
	&=I(\bm{X}_2;W_1W_2) \nonumber  \\
	&=H(W_1W_2)-H(W_1W_2|\bm{X}_2)  \label{X2-express1-mss}
	\end{align}
	and
	\begin{align}
	H(\bm{X}_3)&=I(\bm{X}_3;W_1W_2|W_3)+H(\bm{X}_3|W_1W_2W_3)  \nonumber \\
	&=I(\bm{X}_3;W_1W_2|W_3) \nonumber \\
	&=H(W_1W_2|W_3)-H(W_1W_2|W_3\bm{X}_3).  \label{X3-express1-mss}
	\end{align}
	Next, we use \eqref{X2-express1-mss} and \eqref{X3-express1-mss} to show the bounds in \eqref{rate-mss32-1}-\eqref{rate-mss32-4} one by one.
	
	\subsection{Proof of $2R_i+R_j\geq H(X_2)+H(X_3)$} 
	Due to symmetry, we only need to show this for $i=1$ and $j=2$ which is $2R_1+R_2\geq H(X_2)+H(X_3)$.
	\begin{align}
	&n\cdot\big[H(X_2)+H(X_3)\big]  \nonumber \\
	&=H(\bm{X}_2)+H(\bm{X}_3)  \nonumber \\
	&=\bigg[H(W_1W_2)-H(W_1W_2|\bm{X}_2)\bigg]+\bigg[H(W_1W_2|W_3)-H(W_1W_2|W_3\bm{X}_3)\bigg]   \label{bound-2R1+R2-first}\\
	&=H(W_1W_2)-\bigg[H(W_1W_2|\bm{X}_2)-H(W_1W_2|W_3)\bigg]-H(W_1W_2|W_3\bm{X}_3) \label{bound-2R1+R2-1} \\
	&=H(W_1W_2)+I(W_1;W_2)-\bigg[I(W_1W_2;W_3)+I(W_1W_3;W_2|\bm{X}_2)-I(W_2;W_3|W_1)-I(W_1W_3;\bm{X}_2|W_2)\bigg]  \nonumber \\
	&\quad -H(W_1W_2|W_3\bm{X}_3)  \label{bound-2R1+R2-2}  \\
	&=\bigg[H(W_1)+H(W_2)\bigg]-\bigg[I(W_1W_2;W_3)+I(W_1W_3;W_2|\bm{X}_2)-I(W_2;W_3|W_1)\bigg] +I(W_1W_3;\bm{X}_2|W_2)  \nonumber \\
	&\quad -H(W_1W_2|W_3\bm{X}_3) \nonumber \\
	&=\bigg[H(W_1)+H(W_2)\bigg]-\bigg[I(W_1W_2;W_3)+I(W_1W_3;W_2|\bm{X}_2)-I(W_2;W_3|W_1)\bigg]  \nonumber \\
	&\quad +\bigg[H(W_1)-H(W_1|W_2\bm{X}_2)+I(W_3;W_2\bm{X}_2|W_1)-I(W_1W_3;W_2)\bigg]-H(W_1W_2|W_3\bm{X}_3) \nonumber \\
	&=\bigg[2H(W_1)+H(W_2)\bigg]-\bigg[I(W_1W_2;W_3)+I(W_1W_3;W_2)-I(W_1;\bm{X}_2|W_2)-I(W_3;W_2\bm{X}_2|W_1)\bigg]  \nonumber \\
	&\quad -\bigg[H(W_1|W_2)-I(W_2;W_3|W_1)\bigg]-\bigg[I(W_1W_3;W_2|\bm{X}_2)+H(W_1W_2|W_3\bm{X}_3)\bigg] \nonumber \\
	&=\bigg[2H(W_1)+H(W_2)\bigg]-\bigg[I(W_2;W_3)+I(W_1;W_2)+I(W_1;W_3|W_2\bm{X}_2)\bigg]  \nonumber \\
	&\quad -\bigg[H(W_1|W_2)-I(W_2;W_3|W_1)\bigg]-\bigg[I(W_1W_3;W_2|\bm{X}_2)+H(W_1W_2|W_3\bm{X}_3)\bigg] \label{bound-2R1+R2-3} \\
	&=\bigg[2H(W_1)+H(W_2)\bigg]-\bigg[I(W_2;W_3)+I(W_1;W_2)+H(W_1|W_2)-I(W_2;W_3|W_1)\bigg]  \nonumber \\
	&\quad -\bigg[I(W_1;W_3|W_2\bm{X}_2)+I(W_1W_3;W_2|\bm{X}_2)+H(W_1W_2|W_3\bm{X}_3)\bigg] \nonumber \\
	&=\bigg[2H(W_1)+H(W_2)\bigg]-\bigg[H(W_1|W_2W_3)+I(W_1;W_2)+I(W_1;W_3)\bigg]  \nonumber \\
	&\quad -\bigg[I(W_1;W_3|W_2\bm{X}_2)+I(W_1W_3;W_2|\bm{X}_2)+H(W_1W_2|W_3\bm{X}_3)\bigg]   \label{bound-2R1+R2-4}  \\
	&\leq 2H(W_1)+H(W_2)  \label{bound-2R1+R2-last*}\\
	&\leq n\cdot(2R_1+R_2+3\epsilon) \label{bound-2R1+R2-last}
	\end{align}
	where \eqref{bound-2R1+R2-2} is obtained by
 \begin{align}
 &H(W_1W_2|\bm{X}_2)-H(W_1W_2|W_3)  \nonumber \\
 &=H(W_1W_2|\bm{X}_2)-H(W_1W_2|W_3\bm{X}_2)-I(W_1W_2;\bm{X}_2|W_3)  \nonumber \\
 &=I(W_1W_2;W_3|\bm{X}_2)-I(W_1W_2;\bm{X}_2|W_3) \nonumber \\
 &=I(W_1W_2;W_3)-I(W_1W_2;\bm{X}_2)  \nonumber \\
 &=I(W_1W_2;W_3)-H(\bm{X}_2)  \nonumber \\
 &=I(W_1W_2;W_3)-I(W_1W_3;\bm{X}_2)  \nonumber \\
 &=I(W_1W_2;W_3)-\big[I(W_1W_3;\bm{X}_2W_2)-I(W_1W_3;W_2|\bm{X}_2)\big]  \nonumber \\
 &=I(W_1W_2;W_3)-\big[I(W_1W_3;W_2)+I(W_1W_3;\bm{X}_2W_2)-I(W_1W_3;W_2|\bm{X}_2)\big]  \nonumber \\
 &=I(W_1W_2;W_3)-\big[I(W_1;W_2)+I(W_2;W_3|W_1)+I(W_1W_3;\bm{X}_2W_2)-I(W_1W_3;W_2|\bm{X}_2)\big]  \nonumber \\
 \end{align}
	equality \eqref{bound-2R1+R2-3} is from 
	\begin{align}
	&I(W_1W_2;W_3)+I(W_1W_3;W_2)-I(W_1;\bm{X}_2|W_2)-I(W_3;W_2\bm{X}_2|W_1) \nonumber \\
	&=I(W_1W_2;W_3)+I(W_1W_3;W_2)-I(W_1;\bm{X}_2|W_2)-\bigg[I(W_3;W_1W_2\bm{X}_2)-I(W_1;W_3)\bigg] \nonumber \\
	&=I(W_1W_2;W_3)+I(W_1W_3;W_2)-I(W_1;\bm{X}_2|W_2)-\bigg[I(W_3;W_1W_2)+I(W_3;\bm{X}_2|W_1W_2)-I(W_1;W_3)\bigg] \nonumber \\
	&=I(W_1;W_3)+I(W_1W_3;W_2)-I(W_1;\bm{X}_2|W_2)  \nonumber \\
	&=I(W_1;W_3)+I(W_2;W_3)+I(W_1;W_2|W_3)-I(W_1;\bm{X}_2|W_2)  \nonumber \\
	&=I(W_2;W_3)+I(W_1;W_2W_3)-I(W_1;\bm{X}_2|W_2)  \nonumber \\
	&=I(W_2;W_3)+I(W_1;W_2W_3\bm{X}_2)-I(W_1;\bm{X}_2|W_2)  \nonumber \\
	&=I(W_2;W_3)+I(W_1;W_2)+I(W_1;W_3\bm{X}_2|W_2)-I(W_1;\bm{X}_2|W_2)  \nonumber \\
	&=I(W_2;W_3)+I(W_1;W_2)+I(W_1;W_3|W_2\bm{X}_2),  \label{bound-2R1+R2-additional-last}
	\end{align}
	and the inequality \eqref{bound-2R1+R2-last*} follows from the nonnegativity of Shannon's information measures. Divide both sides of \eqref{bound-2R1+R2-last} by $n$ and let $\epsilon\rightarrow 0$, we can obtain the desired bound
	\begin{equation}
	2R_1+R_2\geq H(X_2)+H(X_3).
	\end{equation}
	
	\subsection{Proof of $R_i+R_j\geq m$} 
	Due to symmetry, we only need to show this for $i=1$ and $j=2$ which is $R_1+R_2\geq m$. Recall that
	\begin{equation}
	m=\max\big\{H(X_2),~\frac{1}{2}H(X_2)+H(X_3)\big\}.
	\end{equation}
	Firstly, it is easy to show that
	\begin{align}
	n\cdot(R_1+R_2+2\epsilon)&\geq H(M_1)+H(M_2) \\
	&\geq H(M_1,M_2) \\ 
	&\geq H(\bm{X}_2) \\
	&=n\cdot H(X_2).  \label{bound-R1+R2-1last}
	\end{align}
	To show the other bound, we consider the following.
	\begin{align}
	&n\cdot\big[\frac{1}{2}H(X_2)+H(X_3)\big]  \nonumber \\
	&=\frac{1}{2}H(\bm{X}_2)+H(\bm{X}_3)  \nonumber \\
	&=\frac{1}{2}\bigg[H(W_1W_2)-H(W_1W_2|\bm{X}_2)\bigg]+\bigg[H(W_1W_2|W_3)-H(W_1W_2|W_3\bm{X}_3)\bigg]   \label{bound-R1+R2-2-1} \\
	&=\frac{1}{2}H(W_1W_2)-\frac{1}{2}\bigg[H(W_1W_2|\bm{X}_2)-H(W_1W_2|W_3)\bigg]+\frac{1}{2}H(W_1W_2|W_3)-H(W_1W_2|W_3\bm{X}_3) \nonumber \\
	&=\frac{1}{2}H(W_1W_2)-\frac{1}{2}\bigg[I(W_1W_2;W_3)+I(W_1W_3;W_2|\bm{X}_2)-I(W_1;W_2)-I(W_2;W_3|W_1)-I(W_1W_3;\bm{X}_2|W_2)\bigg]  \nonumber \\
	&\quad +\frac{1}{2}H(W_1W_2|W_3)-H(W_1W_2|W_3\bm{X}_3) \label{bound-R1+R2-2-3} \\
	&=\frac{1}{2}\bigg[H(W_1)+H(W_2)\bigg]+\frac{1}{2}\bigg[I(W_2;W_3|W_1)+I(W_1W_3;\bm{X}_2|W_2)+H(W_1W_2|W_3)\bigg]  \nonumber \\
	&\quad -\bigg[\frac{1}{2}I(W_1W_2;W_3)+H(W_1W_2|W_3\bm{X}_3)+\frac{1}{2}I(W_1W_3;W_2|\bm{X}_2)\bigg] \nonumber \\
	&=\bigg[H(W_1)+H(W_2)\bigg]+\frac{1}{2}\bigg[I(W_1;\bm{X}_2|W_2)+I(W_3;W_2\bm{X}_2|W_1)-I(W_1W_3;W_2)\bigg] \nonumber \\
	&\quad -\frac{1}{2}\bigg[I(W_1W_2;W_3)+2H(W_1W_2|W_3\bm{X}_3)+I(W_1W_3;W_2|\bm{X}_2)+I(W_1;W_3)\bigg] \label{bound-R1+R2-2-5} \\
	&=\bigg[H(W_1)+H(W_2)\bigg]-\frac{1}{2}\bigg[I(W_1W_2;W_3)+I(W_1W_3;W_2)-I(W_1;\bm{X}_2|W_2)-I(W_3;W_2\bm{X}_2|W_1)\bigg] \nonumber \\
	&\quad -\frac{1}{2}\bigg[2H(W_1W_2|W_3\bm{X}_3)+I(W_1W_3;W_2|\bm{X}_2)+I(W_1;W_3)\bigg] \nonumber \\
	&=\bigg[H(W_1)+H(W_2)\bigg]-\frac{1}{2}\bigg[I(W_1;W_3|W_2\bm{X}_2)+I(W_1;W_2)+I(W_2;W_3)\bigg] \nonumber \\
	&\quad -\frac{1}{2}\bigg[2H(W_1W_2|W_3\bm{X}_3)+I(W_1W_3;W_2|\bm{X}_2)+I(W_1;W_3)\bigg]   \label{bound-R1+R2-2-6} \\
	&\leq H(W_1)+H(W_2)  \label{bound-R1+R2-2-last*} \\
	&\leq n\cdot(R_1+R_2+2\epsilon),  \label{bound-R1+R2-2-last}
	\end{align}
	where \eqref{bound-R1+R2-2-3} follows from \eqref{bound-2R1+R2-1}-\eqref{bound-2R1+R2-2}, 
\eqref{bound-R1+R2-2-5} is obtained by 
\begin{align}
    &I(W_2;W_3|W_1)+I(W_1W_3;\bm{X}_2|W_2)+H(W_1W_2|W_3)  \nonumber \\
    &=I(W_1W_2;W_3)-I(W_1;W_3)+I(W_1W_3;\bm{X}_2|W_2)+H(W_1W_2)-I(W_1W_2;W_3)  \nonumber \\
	&=H(W_1W_2) - I(W_1;W_3) + I(W_1W_3;\bm{X}_2|W_2) \nonumber \\
	&=H(W_1W_2) - I(W_1;W_3) + I(W_1W_3;W_2\bm{X}_2)-I(W_1W_3;W_2) \nonumber \\
	&=H(W_1W_2) - I(W_1;W_3) + I(W_1;W_2\bm{X}_2)+I(W_3;W_2\bm{X}_2|W_1)-I(W_1W_3;W_2) \nonumber \\
	&=H(W_1W_2) -I(W_1;W_3) +I(W_1;W_2)+I(W_1;\bm{X}_2|W_2)+I(W_3;W_2\bm{X}_2|W_1)-I(W_1W_3;W_2) \nonumber \\
    &=\bigg[H(W_1)+H(W_2)\bigg] -I(W_1;W_3) +I(W_1;\bm{X}_2|W_2)+I(W_3;W_2\bm{X}_2|W_1)-I(W_1W_3;W_2), 
\end{align}
\eqref{bound-R1+R2-2-6} follows from \eqref{bound-2R1+R2-additional-last} and the inequality \eqref{bound-R1+R2-2-last*} follows from the nonnegativity of Shannon's information measures. Divide both sides of the two inequalities \eqref{bound-R1+R2-1last} and \eqref{bound-R1+R2-2-last} by $n$ and let $\epsilon\rightarrow 0$, we can obtain the desired bound
\begin{equation}
    R_1+R_2\geq \max\{H(X_2),~\frac{1}{2}H(X_2)+H(X_3)\}.  \label{bound-R1+R2-bound}
\end{equation}
	
	\subsection{Proof of $R_1+R_2+R_3\geq \frac{3}{2}H(X_2)+H(X_3)$} 
	This sum-rate bound is the same as the sum-rate bound for the classical SMDC problem (with $H(X_1)=0$) in \cite{yeung97}. We briefly write out the proof in the following.
	\begin{align}
	n\cdot(R_1+R_2+R_3+3\epsilon)&\geq H(W_1)+H(W_2)+H(W_3)  \label{bound-R1+R2+R3-first} \\
	&=\frac{1}{2}\big[H(W_1)+H(W_2)\big]+\frac{1}{2}\big[H(W_1)+H(W_3)\big]+\frac{1}{2}\big[H(W_2)+H(W_3)\big] \nonumber \\
	&\geq \frac{1}{2}\big[H(W_1,W_2)+H(W_1,W_3)+H(W_2,W_3)\big]  \nonumber \\
	&=\frac{3}{2}H(\bm{X}_2)+\frac{1}{2}\big[H(W_1,W_2|\bm{X}_2) +H(W_1,W_3|\bm{X}_2) +H(W_2,W_3|\bm{X}_2)\big]  \nonumber \\
	&\geq \frac{3}{2}H(\bm{X}_2)+H(W_1,W_2,W_3|\bm{X}_2)   \label{bound-R1+R2+R3-Han's} \\
	&\geq \frac{3}{2}H(\bm{X}_2)+H(\bm{X}_3)    \label{bound-R1+R2+R3-last*}\\ 
	&=n\cdot\big[\frac{3}{2}H(X_2)+H(X_3)\big] \label{bound-R1+R2+R3-last}
	\end{align}
	where \eqref{bound-R1+R2+R3-Han's} follows from Han's inequality and the inequality \eqref{bound-R1+R2+R3-last*} follows from
	\begin{align}
	H(W_1,W_2,W_3|\bm{X}_2)&=H(W_1,W_2,W_3,\bm{X}_3|\bm{X}_2)   \label{bound-R1+R2+R3-additional-1} \\
	&=H(\bm{X}_3|\bm{X}_2)+H(W_1,W_2,W_3|\bm{X}_2,\bm{X}_3)   \nonumber \\
	&\geq H(\bm{X}_3)    \label{bound-R1+R2+R3-additional}
	\end{align}
	where \eqref{bound-R1+R2+R3-additional} follows from the independence of the two sources and the nonnegativity of Shannon's information measures. Divide both sides of \eqref{bound-R1+R2+R3-last} by $n$ and let $\epsilon\rightarrow 0$, we obtain the desired bound
	\begin{equation}
	R_1+R_2+R_3\geq \frac{3}{2}H(X_2)+H(X_3). \label{bound-R1+R2+R3-bound}
	\end{equation}
	Note that the proof in \eqref{bound-R1+R2+R3-first}-\eqref{bound-R1+R2+R3-last} only use the reconstruction constraints in \eqref{recover constraint-mss32-2},\eqref{recover constraint-mss32-3} and the security constraints in \eqref{security constraint-mss32} are not used. Thus we obtain the same bound as the classical SMDC.
	
	\subsection{Proof of $2R_i+R_{i\odot 1}+R_{i\odot 2}\geq 2H(X_2)+H(X_3)$} 
	This bound is also the same as the corresponding bound for the classical SMDC problem (with $H(X_1)=0$) in \cite{yeung97}. We briefly write out the proof in the following. Due to symmetry, we only need to show this for $i=1$, which is 
	\begin{equation}
	2R_1+R_2+R_3\geq 2H(X_2)+H(X_3).
	\end{equation}
	To show the bound, consider
	\begin{align}
	n\cdot\big[2R_1+R_2+R_3+4\epsilon\big]&\geq 2H(W_1)+H(W_2)+H(W_3) \nonumber \\
	&=\big[H(W_1)+H(W_2)\big]+\big[H(W_1)+H(W_3)\big] \nonumber \\
	&\geq H(W_1,W_2)+H(W_1,W_3) \nonumber \\
	&=2H(\bm{X}_2)+H(W_1,W_2|\bm{X}_2)+H(W_1,W_3|\bm{X}_2) \nonumber \\
	&\geq 2H(\bm{X}_2)+H(W_1,W_2,W_3|\bm{X}_2) \nonumber \\
	&\geq 2H(\bm{X}_2)+H(\bm{X}_3)  \label{bound-2R1+R2+R3-last*} \\
	&=n\cdot\big[2H(X_2)+H(X_3)\big]  \label{bound-2R1+R2+R3-last}
	\end{align}
	where \eqref{bound-2R1+R2+R3-last*} follows from \eqref{bound-R1+R2+R3-additional-1}-\eqref{bound-R1+R2+R3-additional}. Divide both sides of \eqref{bound-2R1+R2+R3-last} by $n$ and let $\epsilon\rightarrow 0$, we have
	\begin{equation}
	2R_1+R_2+R_3\geq 2H(X_2)+H(X_3). \label{bound-2R1+R2+R3-bound}
	\end{equation}
	
	\section{Proof of \Cref{thm-32-sMDC}}  \label{proof-thm-32-sMDC}
	It is easy to see that every rate triple in $\cR_2^*$ is achievable by superposition coding of $\bm{X}_1$ and $(\bm{X}_2,\bm{X}_3)$. Next, we show the converse of \Cref{thm-32-sMDC}. The constraints \eqref{rate-sMDC32-1}\eqref{rate-sMDC32-3-2}\eqref{rate-sMDC32-4}\eqref{rate-sMDC32-5} are the same as that of the classical SMDC problem. The proofs can be found in \cite{yeung97}. Thus, we only need to prove \eqref{rate-sMDC32-2} and \eqref{rate-sMDC32-3-1}.
	
	From the conditions in \eqref{recover constraint-3,2-1}-\eqref{security constraint-3,2}, we have 
	\begin{align}
	H(\bm{X}_2)&=I(\bm{X}_2;W_1W_2)+H(\bm{X}_2|W_1W_2)=I(\bm{X}_2;W_1W_2) \nonumber  \\
	&=H(W_1W_2)-H(W_1W_2|\bm{X}_2)  \label{X2-express1-sMDC}\\
	&=\bigg[H(W_1W_2|\bm{X}_1)+I(W_1W_2;\bm{X}_1)\bigg]-\bigg[H(W_1W_2|\bm{X}_1\bm{X}_2)+I(W_1W_2;\bm{X}_1|\bm{X}_2)\bigg] \nonumber \\
	&=\bigg[H(W_1W_2|\bm{X}_1)-H(W_1W_2|\bm{X}_1\bm{X}_2)\bigg]+\bigg[I(W_1W_2\bm{X}_2;\bm{X}_1)-I(W_1W_2;\bm{X}_1|\bm{X}_2)\bigg] \nonumber \\
	&=\bigg[H(W_1W_2|\bm{X}_1)-H(W_1W_2|\bm{X}_1\bm{X}_2)\bigg]+I(\bm{X}_1;\bm{X}_2) \nonumber \\
	&=H(W_1W_2|\bm{X}_1)-H(W_1W_2|\bm{X}_1\bm{X}_2) \label{X2-express2-sMDC} \\
	\end{align}
 and
 \begin{align}
	H(\bm{X}_3)&=I(\bm{X}_3;W_1W_2|W_3)+H(\bm{X}_3|W_1W_2W_3)=I(\bm{X}_3;W_1W_2|W_3) \nonumber \\
	&=H(W_1W_2|W_3)-H(W_1W_2|W_3\bm{X}_3)  \label{X3-express1-sMDC} \\
	&=\bigg[H(W_1W_2|W_3\bm{X}_1)+I(W_1W_2;\bm{X}_1|W_3)\bigg]-\bigg[H(W_1W_2|W_3\bm{X}_1\bm{X}_3)+I(W_1W_2;\bm{X}_1|W_3\bm{X}_3)\bigg]  \nonumber \\
	&=\bigg[H(W_1W_2|W_3\bm{X}_1)-H(W_1W_2|W_3\bm{X}_1\bm{X}_3)\bigg]+\bigg[I(W_1W_2;\bm{X}_1|W_3)-I(W_1W_2;\bm{X}_1|W_3\bm{X}_3)\bigg]  \nonumber \\
	&=\bigg[H(W_1W_2|W_3\bm{X}_1)-H(W_1W_2|W_3\bm{X}_1\bm{X}_3)\bigg] \nonumber \\
	&\quad +\bigg[I(W_1W_2W_3;\bm{X}_1)-I(W_1W_2W_3;\bm{X}_1|\bm{X}_3)\bigg]-\bigg[I(W_3;\bm{X}_1)-I(W_3;\bm{X}_1|\bm{X}_3)\bigg]  \nonumber \\
	&=\bigg[H(W_1W_2|W_3\bm{X}_1)-H(W_1W_2|W_3\bm{X}_1\bm{X}_3)\bigg]+I(\bm{X}_1;\bm{X}_3)-\bigg[H(\bm{X}_1)-H(\bm{X}_1|\bm{X}_3)\bigg]  \nonumber \\
	&=H(W_1W_2|W_3\bm{X}_1)-H(W_1W_2|W_3\bm{X}_1\bm{X}_3) \label{X3-express2-sMDC}
	\end{align}
	Comparing \eqref{X2-express2-sMDC}\eqref{X3-express2-sMDC} and \eqref{X2-express1-mss}\eqref{X3-express1-mss}, we can see that \eqref{bound-2R1+R2-first} can be replaced by a conditional version that each term additionally conditions on $\bm{X}_1$. Then we can check that each step between \eqref{bound-2R1+R2-first} and \eqref{bound-2R1+R2-last} can be replaced by a corresponding conditional version that conditions on $\bm{X}_1$. This finally yields
	\begin{align}
	&H(\bm{X}_2)+H(\bm{X}_3)  \nonumber \\
	&=\bigg[H(W_1W_2|\bm{X}_1)-H(W_1W_2|\bm{X}_1,\bm{X}_2)\bigg]+\bigg[H(W_1W_2|W_3,\bm{X}_1)-H(W_1W_2|W_3,\bm{X}_1,\bm{X}_3)\bigg]   \label{sMDC-bound-2R1+R2-first}\\
	&\leq 2H(W_1|\bm{X}_1)+H(W_2|\bm{X}_1).  \label{sMDC-bound-2R1+R2-first*}
	\end{align}
	Then we have	
	\begin{align}
	&n\cdot\big[3H(X_1)+H(X_2)+H(X_3)\big]  \nonumber \\
	&=3H(\bm{X}_1)+H(\bm{X}_2)+H(\bm{X}_3)  \nonumber \\
	&\leq 3H(\bm{X}_1)+2H(W_1|\bm{X}_1)+H(W_2|\bm{X}_1) \label{sMDC-bound-2R1+R2-2} \\
	&=2H(W_1)+H(W_2)  \label{sMDC-bound-2R1+R2-last*} \\
	&\leq n\cdot(2R_1+R_2+3\epsilon)   \label{sMDC-bound-2R1+R2-last}
	\end{align}
	where \eqref{sMDC-bound-2R1+R2-2} follows from \eqref{sMDC-bound-2R1+R2-first*}. Divide both sides of \eqref{sMDC-bound-2R1+R2-last} by $n$ and let $\epsilon\rightarrow0$, we have 
	\begin{equation}
	2R_1+R_2\geq 3H(X_1)+H(X_2)+H(X_3).
	\end{equation}
	Due to symmetry, we can prove \eqref{rate-sMDC32-2}. Similarly, if we replace \eqref{bound-R1+R2-2-1}-\eqref{bound-R1+R2-2-last} by their conditional versions that condition on $\bm{X}_1$, we can prove \eqref{rate-sMDC32-3-1}. Therefore, \Cref{thm-32-sMDC} is proved.

	%
	
	\section{}  \label{proof-indep-2}
	We state the following lemma which will be used in the proof of \Cref{thm-min_sum_rate-mss}.
	\begin{lemma}
		For $n\geq 1$, let $A_1,A_2,\cdots,A_n, B_1,B_2,\cdots,B_n$, and $C$ be uniform random variables taking values in some finite field $\mathbb{F}_{q}$. All the random variables are mutually independent. If $D_i=A_i+B_i$ for $i=1,2,\cdots,n$, then 
		\begin{equation}
		H(A_1,A_2,\cdots,A_n|C,D_1,D_2,\cdots,D_n)=H(A_1,A_2,\cdots,A_n).
		\end{equation}
	\end{lemma}
	
	\begin{proof}
		Consider the following.
		\begin{align}
		&H(A_1,A_2,\cdots,A_n|C,D_1,D_2,\cdots,D_n)  \nonumber \\
		&=H(A_1,A_2,\cdots,A_n|C)-I(A_1,A_2,\cdots,A_n;D_1,D_2,\cdots,D_n|C)  \nonumber \\
		&=H(A_1,A_2,\cdots,A_n)-I(A_1,A_2,\cdots,A_n;D_1,D_2,\cdots,D_n)  \nonumber \\
		&=H(A_1,A_2,\cdots,A_n)  \nonumber \\ 
		&\quad -\big[H(A_1,A_2,\cdots,A_n)+H(D_1,D_2,\cdots,D_n)-H(A_1,\cdots,A_n,B_1,\cdots,B_n)\big]  \nonumber \\
		&=H(A_1,A_2,\cdots,A_n)  \nonumber \\ 
		&\quad -\big[H(A_1,A_2,\cdots,A_n)+H(D_1,D_2,\cdots,D_n)-H(A_1,\cdots,A_n)-H(B_1,\cdots,B_n)\big]  \nonumber \\
		&=H(A_1,A_2,\cdots,A_n)-\sum_{i=1}^{n}\big[H(D_i)-H(B_i)\big]  \nonumber \\
		&=H(A_1,A_2,\cdots,A_n),
		\end{align}
		where the last step follows from the uniform distribution of $A_i$ and $B_i$ for $i=1,2,\cdots,n$.
	\end{proof}
	
\end{appendices}

\bibliographystyle{ieeetr}
\bibliography{guotao-sMDC_ref}

\end{document}